\documentclass[useAMS, graphicx,twocolumn]{mn2e}

\usepackage{epsf} 
\usepackage{amsmath} 
\usepackage{amsfonts} 
\usepackage{amssymb}
\usepackage{epic}
\usepackage{graphicx}
\usepackage{epsfig}
\usepackage{rotating}
\usepackage{makeidx}
\usepackage{minitoc}
\usepackage{color}
\newcommand{\myfm}[1]{\mbox{$#1$}}
\def\spose#1{\hbox to 0pt{#1\hss}}	
\def\ltabout{\mathrel{\spose{\lower 3pt\hbox{$\mathchar"218$}} 
     \raise 2.0pt\hbox{$\mathchar"13C$}}}
\def\gtabout{\mathrel{\spose{\lower 3pt\hbox{$\mathchar"218$}}
     \raise 2.0pt\hbox{$\mathchar"13E$}}}

\newcommand{\unit}[1]{\ifmmode \:\mbox{\rm #1}\else \mbox{#1}\fi}
\newcommand{\ze}{\ifmmode \mbox{z=0}\else \mbox{$z=0$ }\fi }

\newcommand{\boldv}[1]{\ifmmode \mbox{\boldmath $ #1$} \else 
 \mbox{\boldmath $#1$} \fi}
\renewcommand{\sb}[1]{_{\rm #1}}%

\newcommand{\half}{\myfm{\frac{1}{2}}}

\def\nb1{{\sf NBODY1} }

\newcommand{\tff}{t\sb{ff}}

\newcommand{\rmd}{\ifmmode \:\mbox{{\rm d}}\else \mbox{ d}\fi }
\newcommand{\rmD}{\ifmmode \:\mbox{{\rm D}}\else \mbox{ D}\fi }

\begin{document}

\title{On the equilibrium morphology of systems drawn from spherical collapse experiments}
\date{2005 April 15}
\author[Boily \& Athanassoula]{C.~M. Boily$^{1,\dagger}$ \&  E. Athanassoula$^{2}$
\\          
$^{1}$ Observatoire astronomique de Strasbourg, 11 rue de l'universit\'e, 67000 Strasbourg, France  
\\
$^{2}$Observatoire de Marseille, 2 Place Le Verrier, 13248 Marseille C\'edex 4, France   
\\ 
$^{\dagger}$Formerly at: Astronomisches Rechen-Institut, M\"onchhofstrasse 12-14 Heidelberg, 
D-69120 Germany}

\maketitle

\begin{abstract} We present a purely theoretical study of 
 the morphological evolution of self-gravitating systems 
formed through the dissipation-less collapse of $N$ point sources. 
 We explore the effects of resolution in mass  and 
length on the growth of triaxial structures formed by an instability triggered by an excess of 
 radial orbits. 
 We point out  that as resolution increases, the equilibria shift, from mildly prolate, to oblate. 
A number of particles 
$N \simeq 100,000$ or larger  is required for convergence of axial aspect 
ratios. An upper bound for the softening, $\epsilon \approx 1/256$,  is also identified. 
We then study the properties of a set of equilibria 
formed from scale-free cold initial mass distributions, $\rho \propto r^{-\gamma}$; $0 \le \gamma \le 2$.  
Oblateness is enhanced for initially more peaked  structures (larger $\gamma$'s).  
We map the run of density in space  and find no evidence for a power-law inner structure when 
$\gamma \le 3/2$ down to  a mass fraction $\ltabout 0.1\%$ of the total. 
However when $3/2 < \gamma \le 2$ the mass profile in equilibrium  is well matched
by a power-law of index $\approx \gamma$ out to a mass fraction $\approx 10\%$. 
We interpret this in terms of less effective violent relaxation for more peaked profiles 
 when more phase mixing takes place at the centre.
We map out the velocity field of the equilibria and note that at small radii the velocity coarse-grained 
distribution function is Maxwellian to a very good approximation. 

We extend our 
study to non-scale-free initial conditions and finite but sub-virial kinetic energy. For cold 
collapses the equilibria are again oblate, as the scale-free models. With 
increasing kinetic energy the equilibria first shift to prolate morphology and then  to 
spherical symmetry. 
\end{abstract} 

\section{Introduction} 
This is a purely theoretical  investigation  into the formation of 
equilibria reached through  a phase of gravitational collapse (Lynden-Bell's [1967] violent relaxation process) using numerical $N$-body calculations. Our motivation 
for this undertaking is drawn both from observations and theory, as discussed first below. A survey of recent work  in that area follows.   

\subsection{Cusps in  triaxial galaxies} HST images of elliptical galaxies have revealed 
cuspy ($\rho\propto r^{-\gamma}$) density profiles down to resolution limits 
(e.g., Lauer et al. 1995; Gebhardt et al. 1996; Laine et al. 2003). 
Fits to their luminosity profiles indicate power-law indices ranging from $\gamma \approx 1/2$ to 
$\gamma \approx 2$. These observations have triggered several studies of the origin and orbital content 
of singularities at the heart of galaxies. 
 The orbital structure  of cuspy triaxial galaxies would harbour a broad range of resonant and near-resonant orbits,  including  also 
chaotic orbits (e.g., Merritt \& Fridmann 1996; Holley-Bockelmann et al. 2001). Whether or not the cusp hosts a massive 
(single or binary) black hole  affects both the orbital families that support the cusp as well as  the overall 
mass profile (Holley-Bockelmann et al. 2002; Poon \& Merritt 2001;  Nakano \& Makino 1999; Merritt \& Cruz 2001). 
Furthermore, the diffusion of chaotic orbits in cuspy potentials may lead to a rapid readjustment of the equilibrium 
(Merritt \& Fridman 1996; Kandrup \& Siopis 2003). Thus the possibility of tracking individual orbits in self-consistent 
potentials can hand diagnostics on the stability of ellipticals and on the demographics of black holes in galaxies. 

In this contribution, we construct cuspy triaxial galaxies 
obtained from the classic initial-conditions problem of violent 
relaxation. In particular we are interested in the numerical resolution of central 
density peaks and, at large radii, the global morphology in equilibrium. 
 Analytic and semi-analytic models of triaxial galaxies can be constructed from distribution functions 
 subject to appropriate constraints 
(see e.g., Dehnen \& Gerhard 1994; Holley-Bockelmann et al. 2001; van de Ven et al. 2003). 
However, we chose the flexible three-dimensional $N$-body approach to explore a range of initial conditions and 
study the time-evolution of the systems. This allows total freedom with regard to the symmetry and boundary conditions 
imposed on the system. Drawbacks include finite spatial and mass resolution imposed by limited computer resources. 
This calls on us to check which simulations parameters 
allow  sufficient accuracy and convergence in the numerics, with a view to develop a full library of non-analytic 
models in the most cost-efficient way. 
 
We divide our study in two parts. This contribution is the first part devoted largely to checks of the numerical setup 
and a parameter survey.   In a forthcoming contribution we will explore the families of orbits   
and derive observables from the equilibria obtained. 
Here, first we define the initial conditions for collapse in terms of an  accretion problem 
and give details of the numerical method  (\S2). 
Drawing from known correlations between equilibrium and initial energy distributions (van Albada 1982; Aguilar \& Merritt 1990; Henriksen \& Widrow 1999) we setup sub-virial scale-free power-law mass profiles ($\propto r^{-\gamma}$; see Eq.[1] below). 
We monitor the convergence  
of physical parameters (axis ratios, density profiles) with particle number, $N$, and linear resolution $\epsilon$ in 
\S3.  We then relate equilibrium properties to the initial conditions 
in terms of mass profile and virial ratio (see \S3 and \S4), with the expectations that the power-index of the inner  density
profile  will  match  $\gamma$ of the initial conditions. The properties of a 
 set  of cuspy density profiles are worked out for $\gamma$ in the range $0 \le \gamma \le 2$, covering the range of power indices derived for observed ellipticals (\S5 and \S6). 
  We conclude  with an extended discussion and possible applications of these results in a cosmological 
context. 
 A brief survey of the literature on the topic of violent relaxation is helpful to set our goals in context.

\subsection{Previous numerical work on violent relaxation} 
In a 
 ground-breaking  $N$-body numerical investigation of violent relaxation, van Albada (1982) showed that the de Vaucouleurs $R^{1/4}$ projected luminosity profile of ellipticals could be understood as the outcome  of gravitational collapse. 
This feature proved  attractive since massive  ellipticals are largely pressure-supported with 
little net rotation (Davies et al. 1983; Binney \& Merrifield 1999), a by-product of violent relaxation (e.g. May \& van Albada 1984; Curir \& Diaferio 1994). Later,  radial orbit instabilities (ROI)  were shown to develop in  
numerical  renditions  of anisotropic systems constructed from 
equilibrium distribution functions  $f(E,J^2)$ 
of energy and square angular momentum  (Merritt \& Aguilar 1985; Barnes, Hut \& Goodman 1986) as well as in the end-results of 
violently relaxed systems (Aguilar \& Merritt 1990; Canizzo \& Hollister 1992). 
The ROI develops in e.g. collapsing (sub-virial) spherical distributions due to large radial bulk motion,  so that   
the initial symmetry is lost and the systems become triaxial in equilibrium. The  aspect ratios 
(of mean $\sim 1:2$ [E5]) attained  in  these studies cover the entire spectrum  of ellipticals. 
 Aguilar \& Meritt (1990) demonstrate that the equilibrium distribution functions $f$ satisfies Antonov's stability to 
radial perturbations $\partial f/\partial E < 0$, where $E$ is the binding energy. 
However, van Albada's low-resolution core-halo equilibria
 makes direct application  of these results to  the inner structure of cuspy de Vaucouleur and HST galaxies  
less than straightforward. 
It was noted that relaxation from less smooth (clumpy)  initial conditions  leads to 
more compact virialised equilibria (e.g. McGlynn 1984; Aguilar \& Merritt 1990; Roy \& Perez 2004) possibly offering 
a way forward\footnote{The same 
holds if a smooth collapsing system has random tangential velocities initially (Hozumi et al. 2000; LeDelliou \& Henriksen 2003).}. In the late 1980's the 
emphasis on galaxy formation shifted to include gas cooling and the transformation of spirals to ellipticals 
through mergers (see e.g. Barnes \& Hernquist 1992). Dissipation-less dynamics with $N$-body calculations 
recast  in the framework of hierarchical structure formation in a CDM cosmogony  was found 
to give rise to equilibria  with steep central cusps ($\gamma \gtabout 1$ : Navarro et al. 1997, 2004; 
Fukushige \& Makino 1997, 2001; Fukushige, Kawai \& Makino 2004; Moore et al. 1998, 2004; Diemand et al. 2005). 
Thus repeated episodes of mass accretion  would seemingly lead to 
equilibria with steeper central cusps.  

Both Aguilar \& Merritt (1990) and  
Canizzo \& Hollister (1992) modelled sub-virial mass in-fall 
 with power-law initial mass profiles distributed spherically. 
Each mass shell collapses to the origin at a different time, and the accretion is continued until the last 
mass shell converges to the origin. They have shown that (1) the ensuing 
equilibria are highly prolate; (2) the aspect ratio (defined from the eigenvectors of the
inertia tensor) increases radially and 
is near unity at large radii; (3) the spherically-averaged density profiles of equilibrium systems 
correlates with the initial profile. These investigations were carried out using multi-polar series expansion  and the 
 TREE (Barnes \& Hut 1986) integrators respectively with $N = 5,000$ and 10,000 mass elements. 
 More recent work by Roy \& Perez (2004) used up to $N = 30,000$ particles.  One important result drawn from this study is that homogeneous initial mass profile lead to equilibria that match better cored globular cluster 
 in equilibrium, a conclusion that lends support to the analysis of LMC clusters of Boily et al. (1999). 

  The current, purely theoretical study must be cast in the context of the interesting, full-fledged parameter survey of Roy \& Perez to highlight differences : 
  
1) Recently we have shown that collapse factors in spherical symmetry are indistinguishable from those 
obtained from non-spherical in-fall when the mass resolution $N \ltabout 10^{5} $ (Boily et al. 2002 $\equiv$ BAK+02). 
 This raises issues with the growth of velocity anisotropies near the time of maximum contraction in   
studies with lower resolution. Aarseth et al. (1988) and 
Hozumi et al. (1996) have argued that the tangential velocity dispersion grows faster than the radial 
component at the bounce. Tangential velocities will arise from the growth  of fragmentation modes 
(McGlynn 1984; Aarseth et al. 1988).  
Clearly the velocity field must develop fully for numerical convergence of the end-product properties. 
The results of BAK+02 would suggest a minimum $N \approx 10^5$ in order to 
discriminate between spherical and non-spherical growth modes, by resolving the dynamics at maximum contraction  
well. We confirm in \S4 that numerical convergence is reached when $N$ falls in  this range of values. 

2) Any cold  isotropic distribution of mass develops clumps at the onset of collapse. 
Roy \& Perez (2004) seeded some of their initial  mass  distributions with homogeneous clumps 
(also spherically symmetric). Such initial conditions are difficult to duplicate for 
non-homogeneous spheres as done here owing to the tidal force of the background potential
which does not define a spherical Roche boundary. 
For that reason, we limit our study to distributions of point sources 
with no internal degrees of freedom. This is only a minor setback however because   
 clumps that form in our simulations during in-fall  grow  self-consistently from Poisson
 (root-n) seeds.  These small clumps merge as in-fall proceeds which 
  lead to the growth of a few  large clumps  just prior to maximum in-fall (see e.g. Fig.~7 of Aarseth, Lin \& Papaloizou 1988). Thus the full 
process 
  is highly {\it an}isotropic {\it despite} the choice of spherically symmetric initial conditions. 

3) Roy \& Perez (2004) considered finite-Q (their parameter $\eta$) initial conditions only; we will show that the limit where  $Q \rightarrow 0$  gives rather different equilibria. In that sense, our survey complements theirs while focusing on more specific initial 
conditions. 

It may be worth commenting that 
while we chose power-law initial conditions to generate power-law equilibria, an expectation drawn from past experiments with initial value problems of this kind, we found 
surprisingly that power-laws are recovered only for a sub-set of the parameter space, 
a result that 
turns on its head the long-held belief that the phase-space structure in equilibrium  correlates strongly with the initial conditions chosen. While correlations are found, 
the end-results do differ significantly from those anticipated. 

All numerical studies using particle-based methods have used the virial ratio $Q$ as a free parameter.  
This is partly justified on the grounds that there  may not exist unique diagnostics 
for the growth of ROI either in terms of a critical $Q$ or a global anisotropy parameter in a spherical equilibrium 
(Palmer \& Papaloizou 1987; Perez et al. 1996; see Merritt 1999). Generally lower values of $Q$ lead to deeper radial in-fall and  
more anisotropic velocity fields in equilibrium (more radial orbits), which favours ROI\footnote{The ROI is a Jeans type of instability developing in 
equilibrium systems. The two-stream instability (in- and out-flow) may provide a more appropriate description of the outbreak of a bar during in-fall; 
see Barnes, Hut \& Goodman (1986).}. 
Henriksen \& Widrow (1997) have shown using a one-dimensional code that orbit-crossing which develops during 
in-fall leads to self-similar patterns which break off through a phase-mixing instability. Rapid phase-mixing  
softens the regime of violent relaxation that then sets in. 
Merral \& Henriksen (2003) argue that this instability will develop more slowly when  $Q > 0$. Consequently the similarity pattern persists longer in those cases.  
We will address this point in a study of three-dimensional `warm' collapses 
by comparing their outcome with those of cold collapses. 
 
\section{Method}
\label{sec:method} 

\subsection{Initial conditions} 

\subsubsection{Cold collapses} 
Similarly to other authors (Aguilar \& Merritt 1990 $\equiv$ A\&M+90; 
Cannizzo \& Hollister 1992 $\equiv$C\&H+92; Henriksen \& Widrow 1997) 
we setup a series of spherically symmetric cold scale-free distributions of mass density 

\begin{equation} \rho_o(r) \propto (r/r_s)^{-\gamma} \label{eq:ICs} \end{equation} 
with $0 \le \gamma < 5/2$. The lower limit  corresponds to homogeneous spheres; the upper bound corresponds to systems with finite 
gravitational binding energy $GM^2/r \propto r^{5-2\gamma}$. 
Notice that all models with $\gamma > 1$ have a 
diverging force field at the centre. A value of $\gamma = 3/2$ was adopted as reference for 
the setup of the numerics and convergence of the parameters. 
 The full range of  models is listed in Table~\ref{tab:IC}. 

\subsubsection{Scaling the velocities with $Q$} 
In order to assess how warm initial conditions influence equilibrium profiles, 
we ran also a few simulations in which the particles did not start from rest. 
Instead, they proceed from  Dehnen (1993)  density profiles and matching velocity field. The density profile of these models is given by 

\begin{equation} \rho(r) = \frac{(3-\gamma)\, M}{4\pi}r^{-\gamma}\frac{r_0}{(r_0+r)^\beta}, \label{eq:Dehnen}\end{equation} 

\noindent
where $\gamma$ fixes the inner power-law index, $ \beta = 4 - \gamma $, $r_0$ and $M$ 
are constants fixing the scales of length and total system mass. Because of 
known correlations between initial conditions and equilibria through incomplete 
relaxation  (cf. \S1.1), we expect initial- and equilibrium power-indices of  the mass 
profiles near the centre to be 
equal. Of all the Dehnen models  the one with $\gamma = 3/2$ gives the 
best match to a de Vaucouleur profile, as the reference power-index value adopted for 
cold collapses.

  We define the system virial ratio of kinetic to gravitational energy, $Q$, as 

\begin{equation} Q \equiv \frac{2M \sigma^2 }{W} \label{eq:Q} \end{equation}
\noindent
where $\sigma$ is the three-dimensional velocity dispersion  (we consider only irrotational 
models). $Q = 1$ for a virialised equilibrium, whereas $Q < 1$ ensures 
that all shells of constant mass  converge to the barycentre of the system. 
Several studies of gravitational collapse 
 assign particle velocities randomly (e.g. McGlynn 1984; A\&M+90; C\&H+92; Boily et al. 1999). 
  In the present study, however,  we  assign velocities from the equilibrium distribution 
	function as follows.  
The gravitational binding energy $W$ is computed directly from (\ref{eq:Dehnen}) while the 
global mean square velocity dispersion $\sigma^2$ is obtained from the Jeans equation in spherical symmetry (Binney \& Tremaine 1987, \S 4.2). 	
 
 We then attribute velocities according to an isotropic Maxwellian velocity
distribution constrained to satisfy locally the first three moments of
the Boltzmann equation. 
 The velocities are then renormalised to  achieve the 
	desired global virial ratio $Q$ in (\ref{eq:Q}). This approach has the advantage that 
the velocities are self-consistent with the mass profile of the system, and follows the 
strategy adopted by Barnes et al. (1986) in their study of the growth of ROI's in equilibrium 
systems. 

\subsection{Choice of integration parameters }

\subsubsection{Choice of units}
  Power-law distributions were all truncated at a fixed radius $r_{t} = 2$. 
  We adopted units such that $G = M = 1$, which, together with $r_t$, 
  sets all scales in the problem. For an initial density profile $\rho(r) = \rho_o (r/r_s)^{-\gamma}$, the scales of length $r_s$ and of density $\rho_o$ satisfy $\rho_o r_s^{\gamma} = (3-\gamma)/4\pi M r_t^{\gamma-3} = 2^{\gamma-3} (3-\gamma)/4\pi  $. 
  
\subsubsection{Time-step and free-fall time} 
The evolution of the $N$-body models was followed on the Marseille Grape-5
systems (Kawai et al. 2000), using a specially adapted TREE-code (Athanassoula 
et al. 1998). The code uses a fixed time-step which we set equal to 
\[ \delta t = \tff / 3000 \] where the nominal free-fall time, $\tff$, is  

\begin{equation} \tff \equiv \sqrt{\frac{3\pi}{32G\langle\rho\rangle}} \label{eq:tcr}\end{equation}
where 
 $\langle\rho\rangle = M/(4\pi r_{t}^3/3)$ is the mean density. 
In computing units $\tff \simeq 3.07(0)$. 

Direct application of the virial 
theorem $(r\simeq r_{t}/2)$ at constant $M$ means that the system crossing time  in equilibrium  $t_{cr} = 2r/\sigma_{1d} =  2/\pi\ \tff \simeq 0.637\ \tff$
will take a value close to half the free-fall time, or 1.95(6) $N$-body time units; 
we will use $t_{cr} = 2 $ $N$-body time units for convenience when discussing the results. The code uses a Plummer smoothing $\epsilon$

\[ \phi(r) \equiv \frac{GM}{\sqrt{ r^2 + \epsilon^2} } \] 
 to avoid divergences due to particle-particle interactions. The nature of the problem 
at hand requires a careful setup  to ensure that global energy and angular 
momentum are preserved to good accuracy throughout evolution. 
The maximum relative error $\delta E/E$ (in percentage) measured during the complete evolution fluctuates between different runs, but remains  of the order of,
or better than, a few parts in a thousand for the adopted values of the opening
angle, softening and time step. We discuss these values   briefly. 

\subsubsection{Choice of softening $\epsilon$}

As in all cases, the value of $\epsilon$ should be tailored to the problem 
at hand. The softening must be large enough to avoid collisions between 
particles, while being small enough to resolve time-variations 
in the potential at all stages of evolution. Our problem is particularly 
difficult, since it includes both very high density and very low density 
regions. Furthermore, particles do not stay all through the evolution in
regions of similar density, but can visit regions of very different density. 

We tried a number of softening lengths ranging from 1/32 to 1/1024 and after many tests, 
some of which 
are described in section~\ref{sec:morphology}, we adopted a reference smoothing length of 
1/512. The effect of changing this value and how this influences results 
are discussed in section~\ref{subsec:softening}. 

Of course we can not trust our results for distances smaller than a few 
times the softening length. 
In practice we took a conservative radius of $5\times \epsilon$ as the 
inner-most radius to compute physical parameters (velocity dispersion, density, etc).

\subsubsection{Choice of opening angle $\theta_c$}

The TREE method defines a critical angle $\theta_c$ to control the 
accuracy of a limited expansion to the force field. The choice of 
$\theta_c$ is a compromise between high accuracy and performance. 
In most simulations an opening angle $\theta_c \approx 0.7$ allows  
energy conservation typically to the order of 0.1\% (Barnes \& 
Hernquist 1996; Athanassoula et al. 2000). 
For collapse calculations, however, as for mergings, the time-variations 
of the  potential are rapid and large, so such an opening angle may not 
necessarily suffice.
 We have thus carried out a series of calculations varying $\theta_c$ and Plummer softening to determine which combination of parameters allows integration to the desired  accuracy. Our results proved robust for a wide range of opening angles (cf. Fig.~\ref{fig:AC}). We adopted a median value $\theta_c = 0.4$ for all our calculations. Simulations with an opening angle of 0.7 give a somewhat 
larger value of the energy variation. We thus did not include
them in the analysis presented in the next sections. In general, 
runs that gave energy errors larger than 
1\%  were discarded from analysis altogether. 

\subsection{Physical description} 
\begin{figure*}
\setlength{\unitlength}{1cm} 
\begin{center}
\begin{picture}(8,11)(0,0) 
	\put(-2.5,3.5){ \epsfysize=.75\textwidth 
		    \epsffile[ 100 150 450 600]
		   {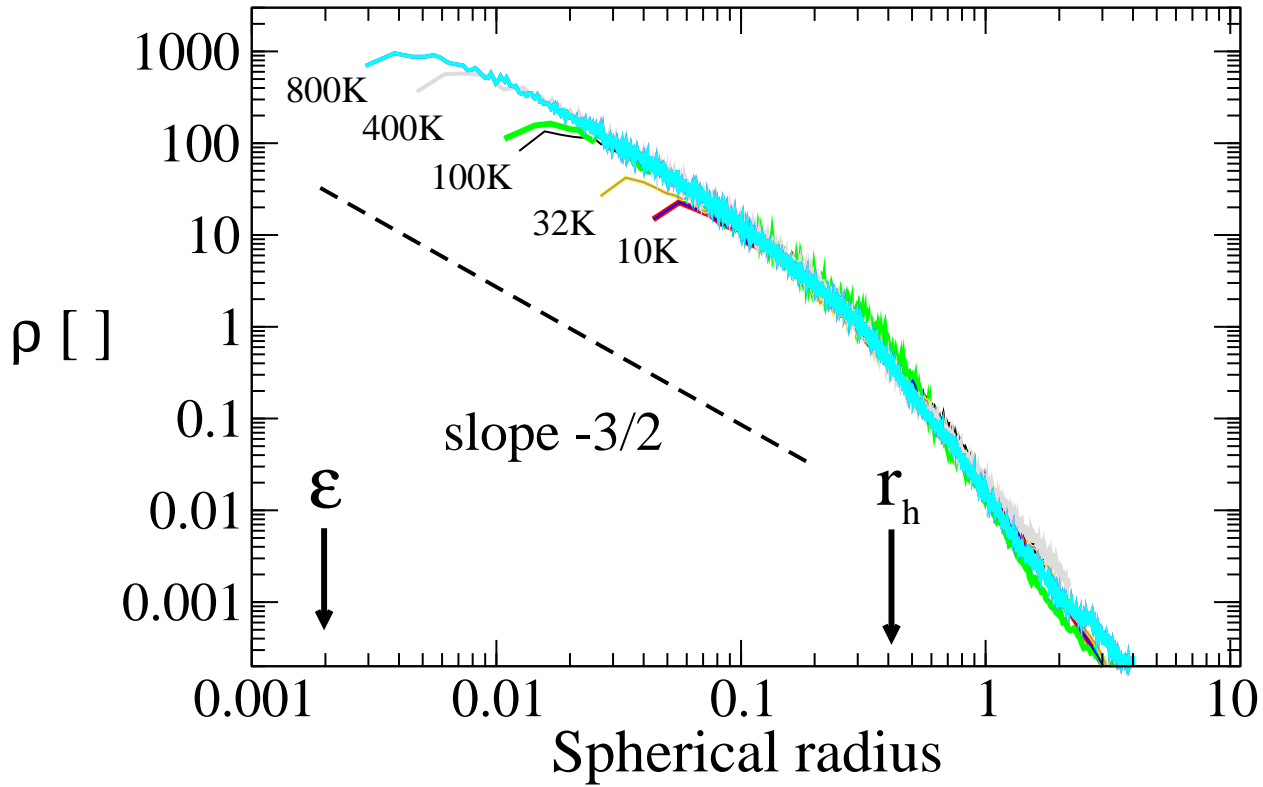} 
                  }
\end{picture} 
      \caption{Density profiles of a series of runs with $\gamma = 3/2$. The number of particles for each 
case is indicated. The density was averaged over spherical shells containing 100 particles 
each, centered on the density maximum identified using the six-nearest neighbour scheme of Casertano \& Hut (1985). The larger simulations allow 
deeper probing of the inner region. 
Two runs with $N = 100,000$ particles but different random number seeds are displayed to illustrate scatter. 
The smoothing length and half-mass radius are marked with arrows, and  a 
straight line of slope -3/2 ($\rho \propto r^{-3/2}$) is shown for reference. } 
         \label{fig:Rhovsr}
\end{center}\end{figure*}

Starting with zero kinetic energy, all particles converge toward the centre 
of gravity which coincides with the centre of coordinates. Since the free-fall time 

\begin{equation} \tff \propto \frac{1}{\sqrt{G\langle\rho\rangle} } \propto (r/r_s)^{\gamma/2}\label{eq:tff} \end{equation}  
increases with radius, several orbits cross at the centre before a significant fraction 
of the mass has reached there. This feature allows us to treat the problem as an 
accretion problem.  At constant radius, the flux of matter through a shell during in-fall 

\begin{equation} \dot{m} = 4\pi r^2 \rho(r,t) v_r(t) \propto r^{3-3\gamma/2} \propto t^{3(2-\gamma)/\gamma} \label{eq:mdot} \end{equation} 
 increases with time when $\gamma \le 2$. The above relation was derived 
 by following an in-falling  shell of constant internal mass and so holds up to first orbit crossing. As orbits begin to cross at the centre, a pattern emerges from the origin and propagates outwards. At a given radius, out-going particles eventually 
 meet with in-falling material and the net mass in-flux 
 $\dot{m}$ drops. Thereafter a self-consistent equilibrium is established 
over some $\approx$ 20 $N$-body units of time (or, $\approx 10\, t_{cr}$); all our simulations ran for 80 $N$-body units of time, or $\approx 40$ dynamical  times, 
to ensure stability of the equilibrium configurations. 
Note that the classical two-body relaxation time at the system half-mass radius, $t_{rh}$, is given by  (see Binney \& Tremaine 1987) 
\[ \frac{t_{rh}}{t_{cr}} \simeq \frac{1}{10} \frac{N}{\ln 0.4 N} \] 
so that for $N = 100,000$ particle calculations  $t_{rh} \approx 900\, t_{cr}$ 
is much larger than the total runtime of the simulations. In fact, the 
{\it central} relaxation time, defined in terms of the dynamical time at the centre, can accommodate an $t_{cr}$ shorter by a factor 20, or 
an increase in density of $20^2 = 400$, and still remain clear of two-body relaxation effects. 
We checked explicitly for several cases that runs with this number of particles did not exceed a density contrast 
 (central to half-mass  values) of 100 (cf. Fig.~\ref{fig:Rhovsr}). 
We have performed a few runs with particle number $N \sim 10^4$, 
to ease comparison with results from previous papers where this value of  $N$ 
was used. Runs with $N \sim 10^4$ have $t_{rh} \sim 160\, t_{cr}$ at the half mass 
radius; the maximum density contrast allowed is now $\simeq 16$, close to what was obtained (Fig.~\ref{fig:Rhovsr}).  These runs will likely have suffered a degree of two-body relaxation at the centre~: to ensure that our analysis was not  affected by relaxation for this case, we also took snapshots at time $t = 40$ (half the evolution time) and 
 verified that very similar conclusions applied. Therefore only results obtained for $t = 80$ units of evolution 
will be presented.  We note that the  relaxation time estimates are conservative 
since they do not take account of softening. This cuts off large-deflection angle encounters and 
lengthens collisional relaxation (see e.g. Theis \& Spurzem 1999; Athanassoula, Vozikis \& Lambert 2001). 


\section{Example : time-evolution of an $\gamma = 3/2$ run} 
\label{sec:example}

\subsection{Centre of density vs centre of mass} 
Gravitational collapse leads to a spread of the individual particle energy range.  Particles  
acquiring  positive energy leave the system on a dynamical time-scale. 
Overall, on the order of 10\% of the particles escape. The  fraction of escapers is a 
function of both accretion index $\gamma$ and initial morphology. 
 The combined effects may cause  up to $\approx 22\%$
  mass loss for homogeneous spherical systems (Table 1 of C\&H+92; Boily 1994). 
As a result of the growth of radial orbit instability and of anisotropic 
loss of unbound particles, the centre of coordinates does not necessarily match the 
centre of mass of the bound particles. We found in most cases a small but non-zero 
linear momentum carried away by escapers. 
 In all the analysis, and in particular when
fitting isodensity curves and calculating the inertia tensors and
their eigenvalues, we recentered the coordinates to the center of
density (i.e. the maximum density point) in order to avoid errors due
to off-centered isophotes.  The density maximum was 
identified using the six-nearest neighbour scheme of Casertano \& Hut (1985).  
 The approach allows a much 
higher resolution of density profile around the centre, down to a few $\times \epsilon$ (Fig.~\ref{fig:Rhovsr}). 
 The NEMO package utility functions greatly helped automate the procedure 
(see http://bima.astro.umd.edu/nemo/; Teuben 1995). 

\subsection{Inside-out morphology} 
\label{subsec:ex_morpho}

We investigated the morphology of equilibrium profiles in the centre of density 
frame in two different ways. First we sorted the particles in 
binding energy and removed the 20\% particles 
with the largest energy (which includes all unbound particles). We then divided the remaining particles in four bins, each containing 20\% of the initial particle 
number. In this manner we did not take into consideration unbound particles and particles 
bound to the system but in a very-low density region: this imitates the effect 
of a  tidal boundary. We also used an alternative approach, 
which consisted in  sorting particles by increasing spherical radius and removing the outermost 20\%. In both cases we re-centered on the density maximum of the remaining particles. 
The two approaches gave similar results.

We  computed the inertia tensor of selected particles, separately 
for each bin. The orientation of the system was set so that Cartesian 
axes matched the principal axes of the bound particles, with the positive x-axis 
coinciding with the semi-major axis. We then defined ellipsoidal rms axes $a > b>c$ 
from the inertia tensor $I$ as follows, for instance for the minor axis $c$  

\[ c^2 = \langle z^2 \rangle = \frac{ I_x + I_y - I_z }{ 2 m N_i } \] 
where the $\langle..\rangle$ is the average of a quantity over $N_i$ particles of mass $m$ 
in the i$^{th}$ bin. 
 From our simulations we  recover  spatially- and time-dependent axial 
ratios. \newline 

\begin{figure}
\setlength{\unitlength}{1cm} 
\begin{center}
\begin{picture}(8,7.2)(0,0) 
	\put(0.5,1.3){ \epsfysize=.4\textwidth 
		    \epsffile[ 100 150 450 600]
		   {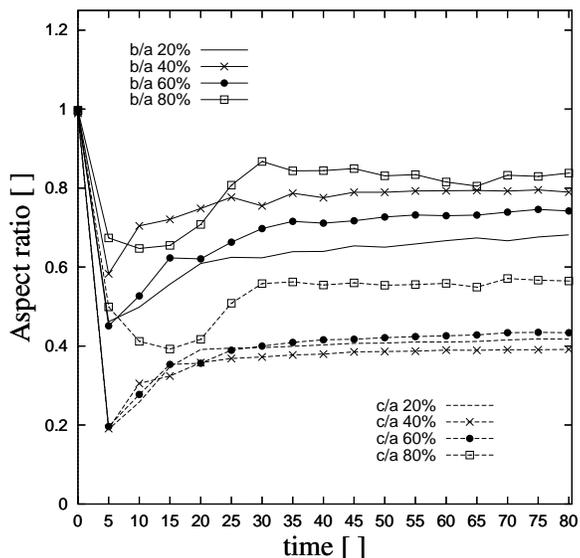} 
                  }
\end{picture} 
      \caption{Time-evolution of the axis ratios for run d007.} 
         \label{fig:Ratio007}
\end{center}\end{figure}

Figure~\ref{fig:Ratio007} graphs the time-evolution of the axis-ratios $b/a$ and $c/a$ for four bins for 
the duration of an $N = 800,000$ run (d007 in Table~\ref{tab:IC}). The configuration 
initially 
had a power index $\gamma = 3/2$. Since we start from spherical symmetry, both axial ratios are initially equal to 1.  
However, the rapid 
growth of radial motion leads to instability and the configuration quickly becomes triaxial (around $t = 5$ on the 
figure). Axial evolution slows down rapidly, until after $t \approx 20 $ when the curves flatten out and begin 
to fluctuate about average values. What morphological 
evolution remains is largely confined to the innermost 20\% mass bin, seen both for the minor c-axis and the 
median b-axis. Thus the inner region continues to evolve, albeit slowly, within the overall relaxed structure. 
 A close look at Fig.~\ref{fig:Ratio007} shows a gentle but steady 
increase of the minor-axis ratio $c/a$, from $\approx 0.38$ to $ \approx 0.43$, for the innermost three mass bins of the system.  The minor-axis ratio of the outer-most 
particles show no indication of evolution. 
By contrast, the ratio $b/a$ evolves significantly for the  
20\% most bound particles only, from $0.58$ to $\approx 0.65$.  
This is suggestive of a trend toward axial symmetry as argued 
by Theis \& Spurzem (1999) and Heller (1999) in their relaxation study of Plummer spheres (see also Curir \& Diaferro 1994) . 
  The example depicted on Fig.~\ref{fig:Ratio007} suggests more rapid evolution in the inner, 
denser region of the system (shortest dynamical time-scale).  We note that the initial conditions used here are without a harmonic (Plummer) core. 
The similar qualitative trends obtained from  distinct sets of initial conditions 
 supports the view that evolution toward axisymmetry is a generic feature of collision-less triaxial  equilibria. 
\newline


\section{Morphology of the equilibrium profiles} 
\label{sec:morphology}

We now turn to a more quantitative inspection of our models. 
A\&M+90 found from their 5,000-particle runs 
 that the bulk of their triaxial equilibria shaped by  
the radial orbit instability 
 are prolate (see their Table~1). This result was 
confirmed  by C\&H+92, who showed using 
10,000-particle runs that the 
morphology of virialised objects  varies 
in space, the equilibria  being more triaxial prolate at the half-mass 
radius, and near-spherical in the inner and outer parts (see their Table 1 and Fig.~4).   
These studies set references against which to compare  our results. 
In this section, we discuss  the 
effect of varying numerical parameters (such as the number of
particles and the softening) and model parameters (such as the index
$\gamma$) on the 
equilibrium morphology; we consider parameterised fits to the density profiles in \S5.


   \begin{figure*}
\setlength{\unitlength}{1cm} 
\begin{picture}(8,8)(0,0) 
	\put(-4.,2.){ \epsfysize=0.4\textwidth
		    \epsffile[ 100 150 450 600]
		   {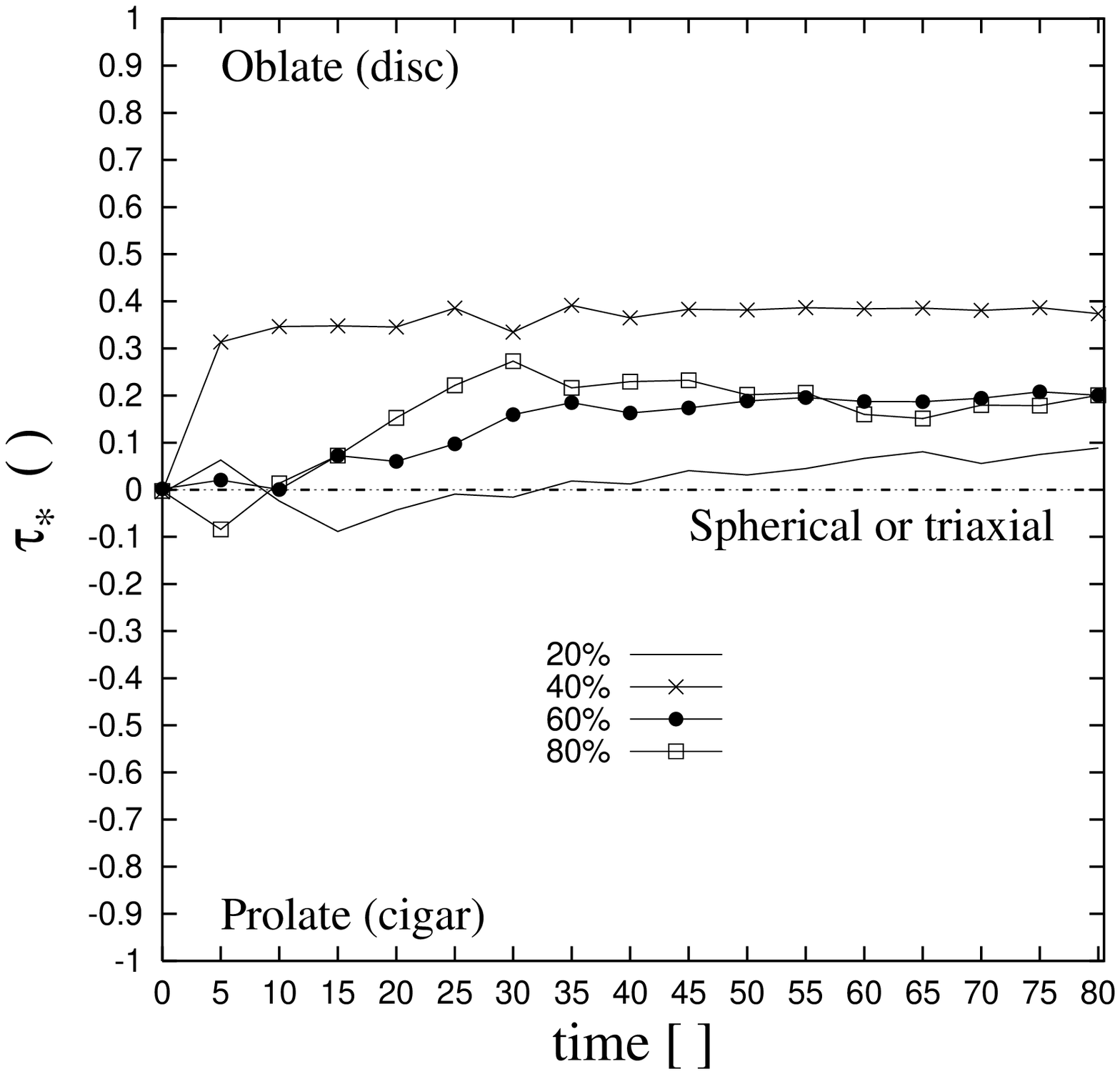} 
                  }
	\put(5.,2.){ \epsfysize=0.4\textwidth
		    \epsffile[ 100 150 450 600]
		   {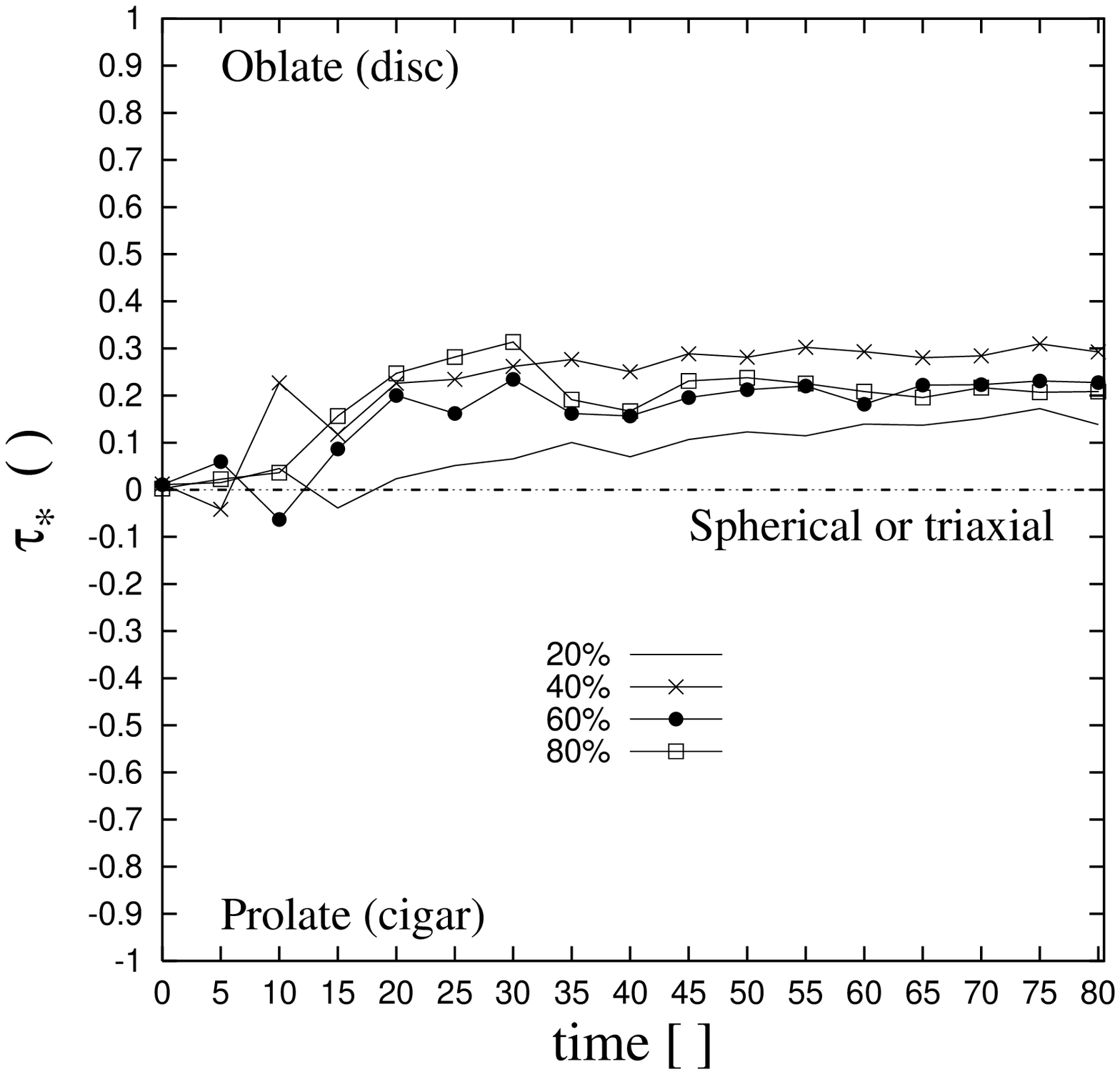} 
                  }
\end{picture} 
      \caption{Time-evolution of runs d007 ($N$ = 800,000 particles, left-hand panel) and d022 
($N$ = 200,000 particles, right-hand panel). We graph the morphological parameter
 $\tau_\ast$ defined in (\ref{eq:taustar}) for different 20\%-mass shells, as indicated. }
         \label{fig:Taustar}
   \end{figure*}

\subsection{Quantifying morphology} A\&M+90 introduced the parameter 

\begin{equation} \tau \equiv \frac{ b - c }{a - c } \label{eq:tau} \end{equation} 
to put the morphology of the relaxed equilibria on a quantitative footing. 
$\tau$ is bounded between 0 (prolate spheroids) and 1 (oblate spheroids). The degree of oblateness or prolateness varies within the bracket $[0,1]$ :  
  configurations with $\tau > \half$ are  oblate, otherwise they are prolate.  
 
In practice $\tau$ is  sensitive to round-off errors when $a \approx b \approx c$. A clearer distinction between the 
three morphologies (triaxial, prolate and oblate) would be desirable. Consider instead 
 the parameter 

\begin{equation} \tau_\star \equiv \frac{b^2- c^2}{b^2+c^2}  - \frac{a^2 - b^2}{a^2 + b^2} 
\label{eq:taustar} \end{equation} 
which covers the range $[-1,1]$;  $\tau_\star = 0$  for a spherical distribution, while $\tau_\star > 0 $ for `oblate' triaxial distributions and $< 0$ for `prolate' triaxial distributions. 
 The parameters $\tau$ and $\tau_\ast$ are related to each other and both 
 increase monotonically with oblateness, however note that  $\tau_\ast$ allows to 
identify axisymmetric discs ($\tau_\ast = 1$) or spindles ($\tau_\ast = -1$) 
unambiguously. For this reason  we have chosen to use $\tau_\ast$ in our analysis. 

As an example, the evolution of $\tau_\ast$ with time
is shown on Fig.~\ref{fig:Taustar}, left-hand panel, for the $\gamma = 3/2$ 
case displayed on Fig.~\ref{fig:Ratio007}.  
Although on the whole  the 
system  achieves oblate morphology rapidly, the aspect ratios and hence the diagnostic $\tau_\ast$ remain 
strong functions of the volume sampled. 
The innermost 20\% mass shell first becomes mildly prolate  before 
 shifting to oblate shaped ($\tau_\ast > 0$) after some 30 time units of evolution.  
This clearly serves as 
warning against hastily classifying the system globally  as either prolate or oblate. 

\subsection{Dependence on $N$} 
We looked for trends in axis ratios $c/a, b/a$, and in $\tau_\ast$ as function of the number of particles used 
in the calculations. Recall that relatively large variations are 
expected for $N < 10^5$ (BAK+02). To show that the results vary little whenever $N > 100,000$, first  we 
graph on the right panel of Fig.~\ref{fig:Taustar} the results of the same calculation as shown on the left but now with $N = 200,000$ 
particles. The trends and values over time of $\tau_\ast$ 
are essentially the same as for the larger $N = 800,000$ calculation.   This suggests that  the numerics have converged 
for that range of values of $N$. However, full convergence has to be demonstrated 
through comparison with  a set of results for smaller particle number. 

We selected eight runs with $N$ ranging from 
$N = 10,000$ to 800,000. Each run 
had a power-index $\gamma = 3/2$. To minimise 
 root-$N$ fluctuations we averaged aspect ratios over time from 60 to 80 (end of the calculations) for a total of 5 outputs. 
Table~\ref{tab:N} lists their simulation parameters and  aspect ratios and $\tau_\ast$ achieved in equilibrium.  
(The aspect ratios were evaluated separately for four 20\%-mass bins and for each output, however only values averaged over all mass bins are given.)  
We added to this compilation the results obtained by C\&H+92 for this value of $\gamma$ (their index $n$) and 
 $N = 10,000$, as well as results from A\&M+90, who considered cases with $\gamma = 1$ and $N = 5,000$. 
(A comparison of results from C\&H+92  for the same $\gamma = 1$  values 
with those of A\&M+90 shows  a  good agreement between these two studies.) 
 The values of $\tau_\ast$ computed from the data 
  of A\&M+90 were increased by +0.15 in account of the fact that they used a smaller 
  $\gamma$ and in anticipation of the trend seen in plots of  $\tau_\ast$ versus
   $\gamma$ (see \S4.4 and Fig.~\ref{fig:Tauvsgamma} below). 

The results are graphed on Fig.~\ref{fig:Tau}, right-hand panel.  
We find predominantly oblate equilibria ($\tau_\ast > 0$) for all our simulations with $N \ge 25,000$.  Fig.~\ref{fig:Tau} indicates less  
oblate structures from the $N=10,000$ cases than
for larger-$N$ runs  (open squares on the figure).
We observe that the {\it full range} of axis ratios is reached only 
for $N \gtabout  100,000 $. Thus we find equilibria with ratios $a/c \approx 2.6$ 
for that order of particle number, well above
the maximum $a/c \approx 2$ for 10,000-particle runs (cf. Fig.\ref{fig:AC}, middle panels). The same conclusion applies to the major-to-median axes ratio. 

The major- to minor-axis ratios of our 10,000 particle run, ranging from 1.6 to 2, all exceed those obtained 
by C\&H+92 for the same $\gamma = 3/2$ case. What is more, the results  
we obtained for $\tau_\ast$ differ significantly from those 
of both A\&M+90 and C\&H+92 (we get $\tau_\ast \approx -0.308$ from their Table~1): 
such results point  systematically to prolate structures of equilibrium. 
Differences between our results and 
those of C\&H+92 can not be attributed  to particle number : below we make a detailed  comparison of the
initial conditions and software configurations to understand the origin of these differences. 
 
 
   \begin{figure*}
\setlength{\unitlength}{1cm} 
\begin{picture}(8,8)(0,0) 
	\put(-3.3,2.5){ \epsfysize=4.20in
		    \epsffile[ 100 150 450 600]
		   {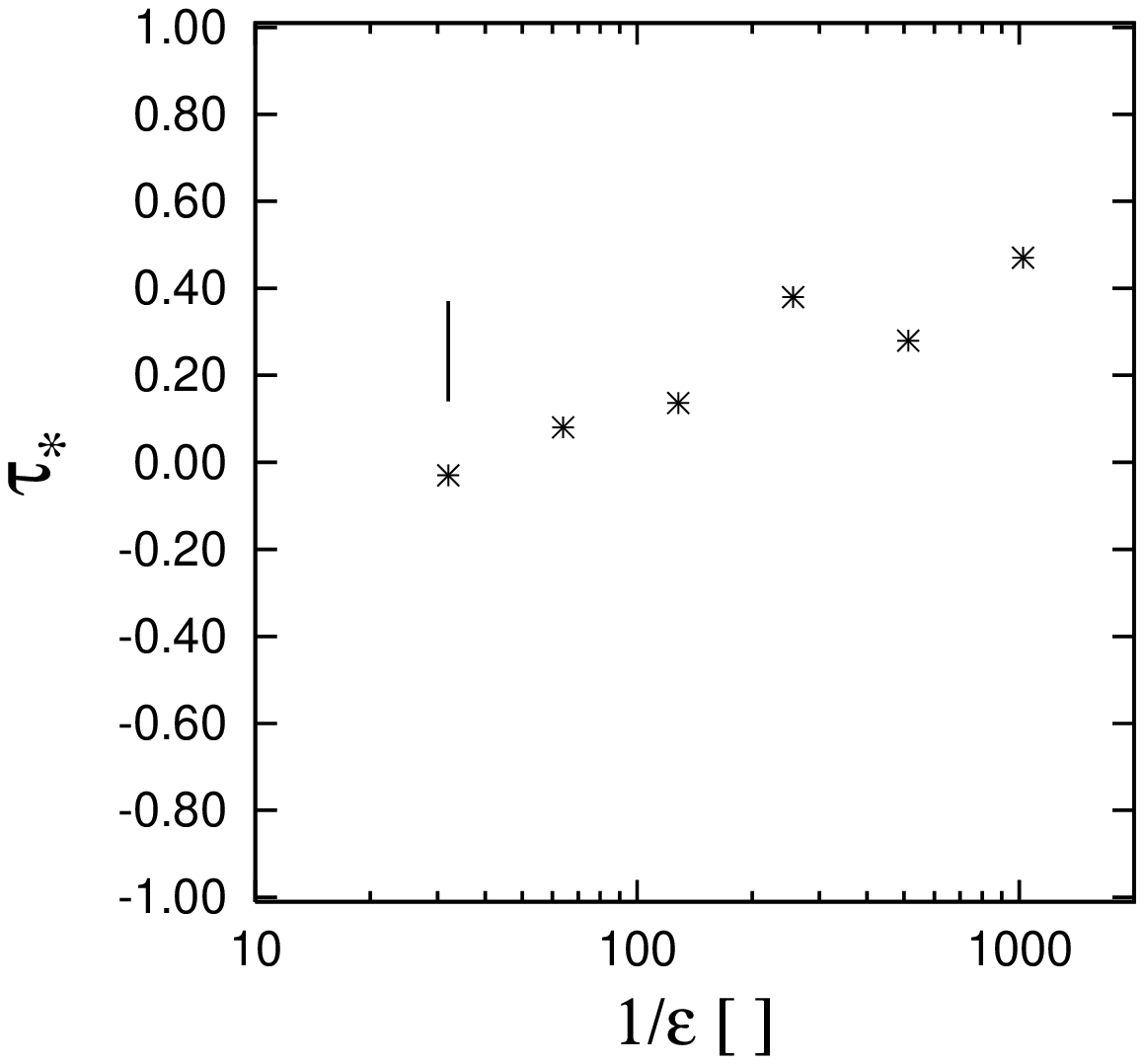} 
			}
		\put(-2.5,3.0){ {\Huge $\dagger$} } 
	\put(5.45,2.5){ \epsfysize=4.20in
		    \epsffile[ 100 150 450 600]
		   {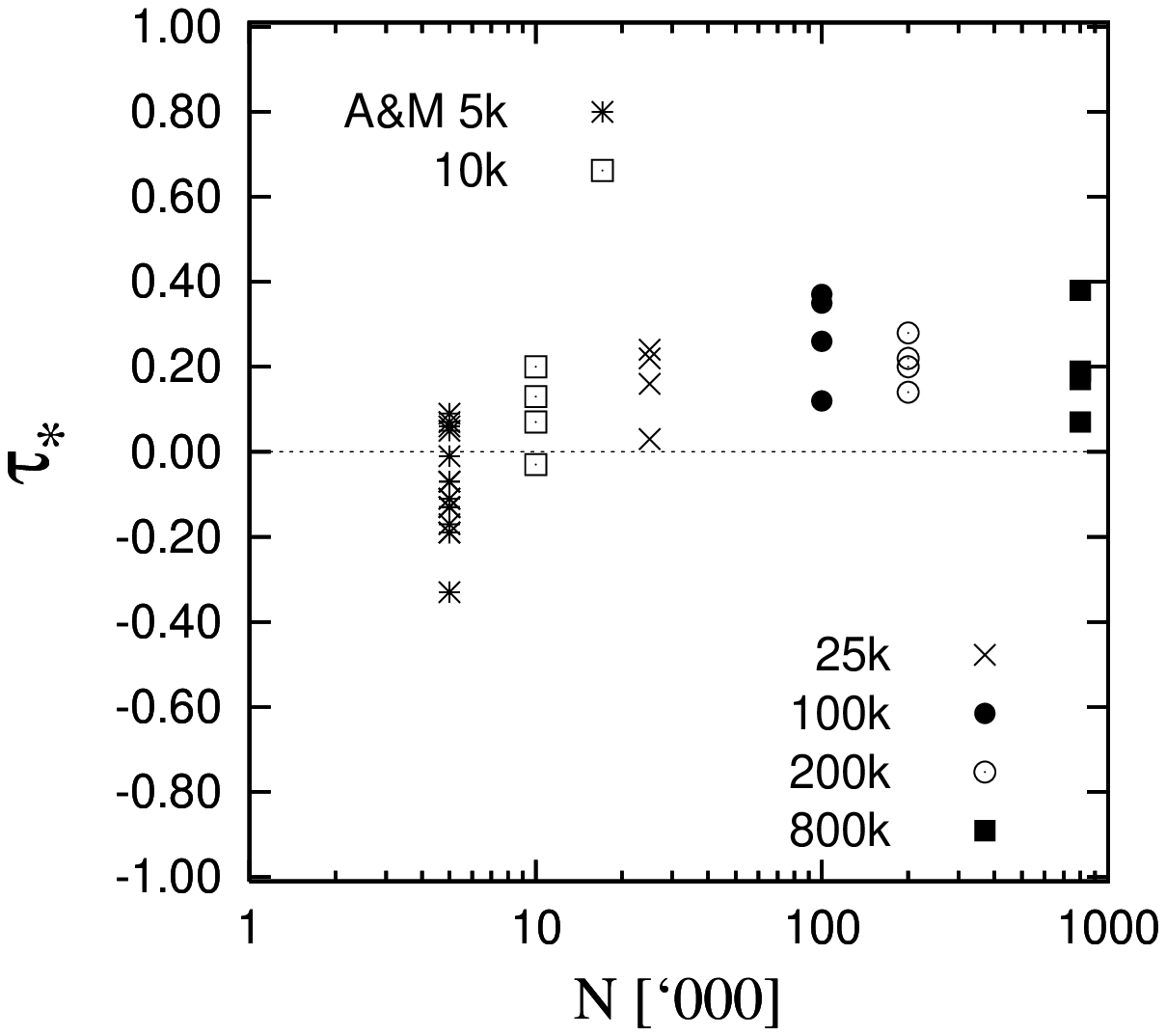} 
}
		\put(8,3.0){ {\Huge $\dagger$} } 
\end{picture} 
      \caption{Global morphology parameter $\tau_\ast$ as function of inverse smoothing length  $1/\epsilon$ (left-hand 
panel) and particle number $N$ (right-hand panel). The vertical tick mark on the left panel 
 is an error estimate derived from the scatter seen in the axial ratios; the data points  
are mean values for the innermost 80\% mass. The data for the 5,000-particle runs are taken 
from A\&M+90 but shifted upwards by +0.15 (see text for details). The symbol `$\dagger$' 
is the mean value lifted from C\&H+92 for $N = 10,000$.} 
         \label{fig:Tau}
   \end{figure*}

   \begin{figure*}
\setlength{\unitlength}{1cm} 
\begin{picture}(8,12)(0,0) 
	\put(-4.5,7.25){ \epsfysize=2.5in
		    \epsffile[ 100 150 450 600]
		   {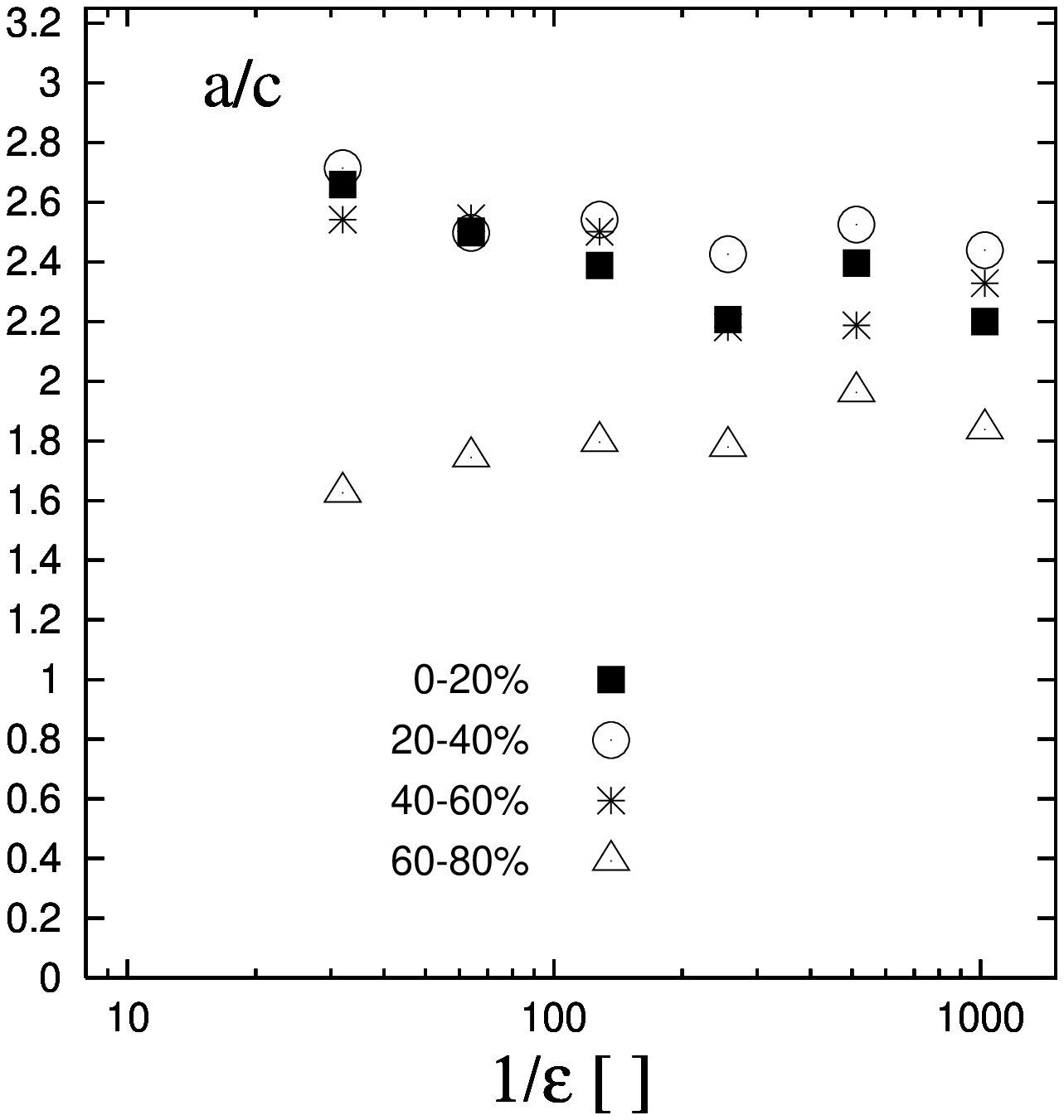} 
}
	\put(-4.5,1.5){ \epsfysize=2.5in
		    \epsffile[ 100 150 450 600]
		   {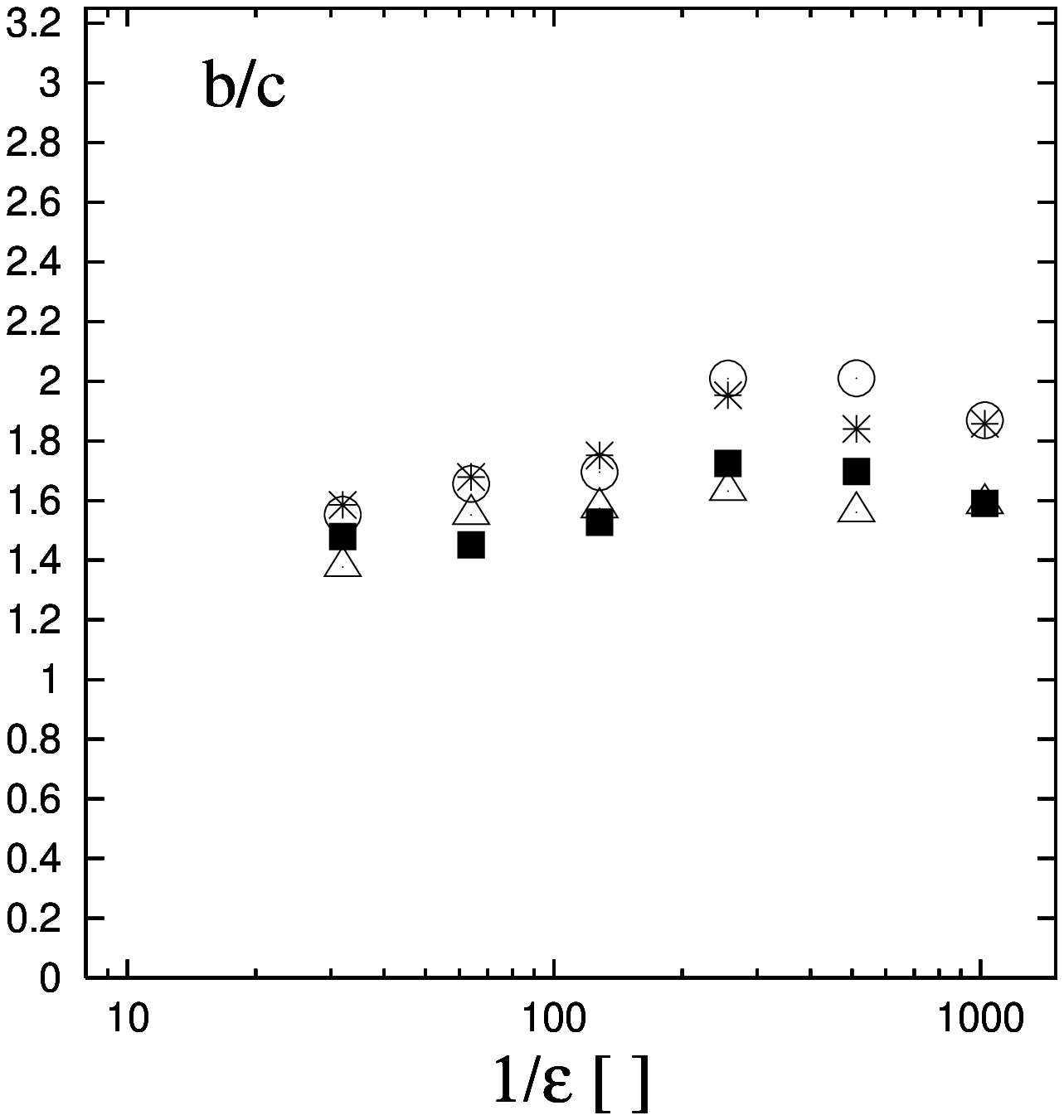} 
                  }	

	\put(7.2,7.25){ \epsfysize=2.50in
		    \epsffile[ 100 150 450 600]
		   {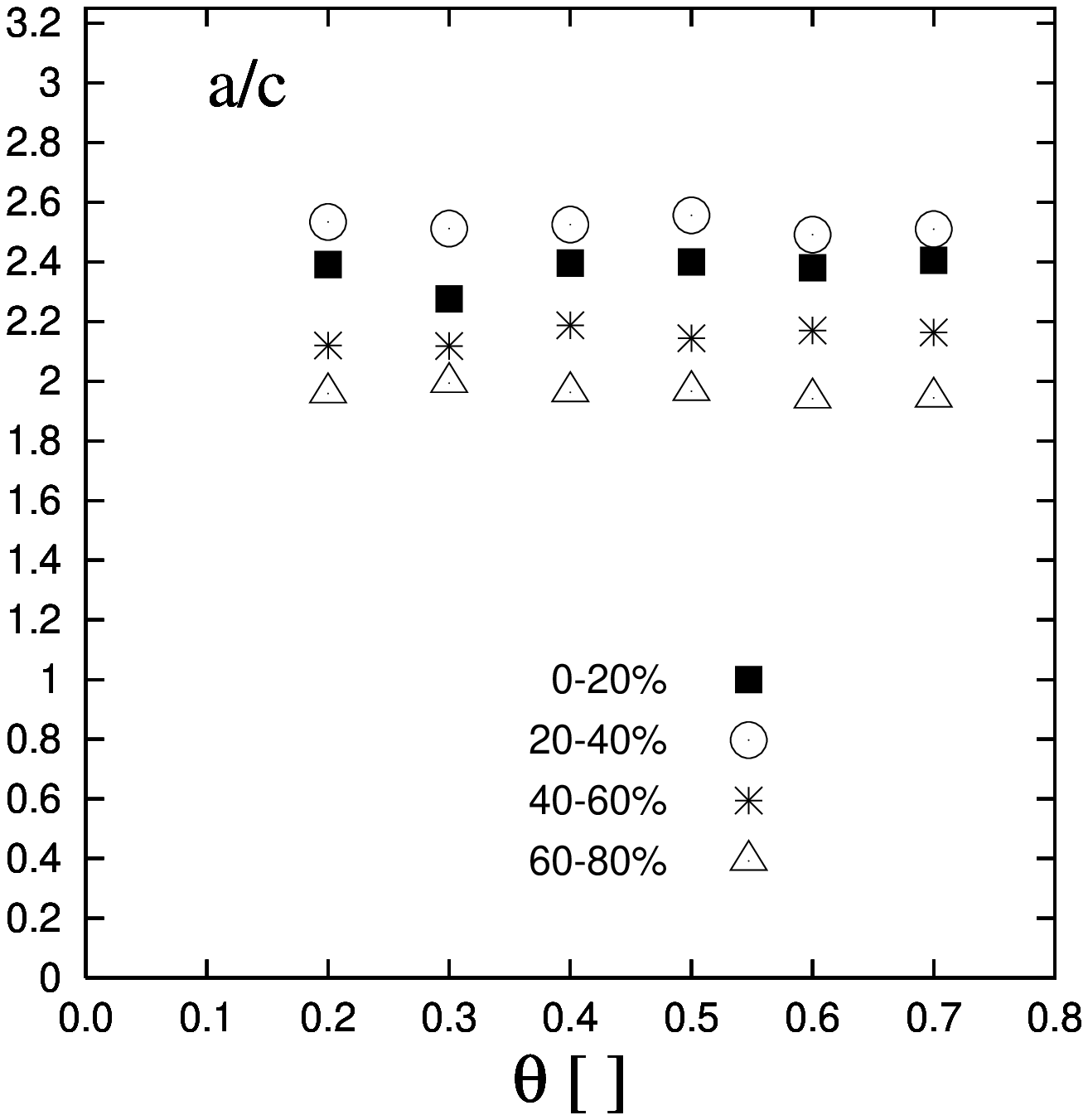} 
			}
	\put(7.2,1.5){ \epsfysize=2.50in
		    \epsffile[ 100 150 450 600]
		   {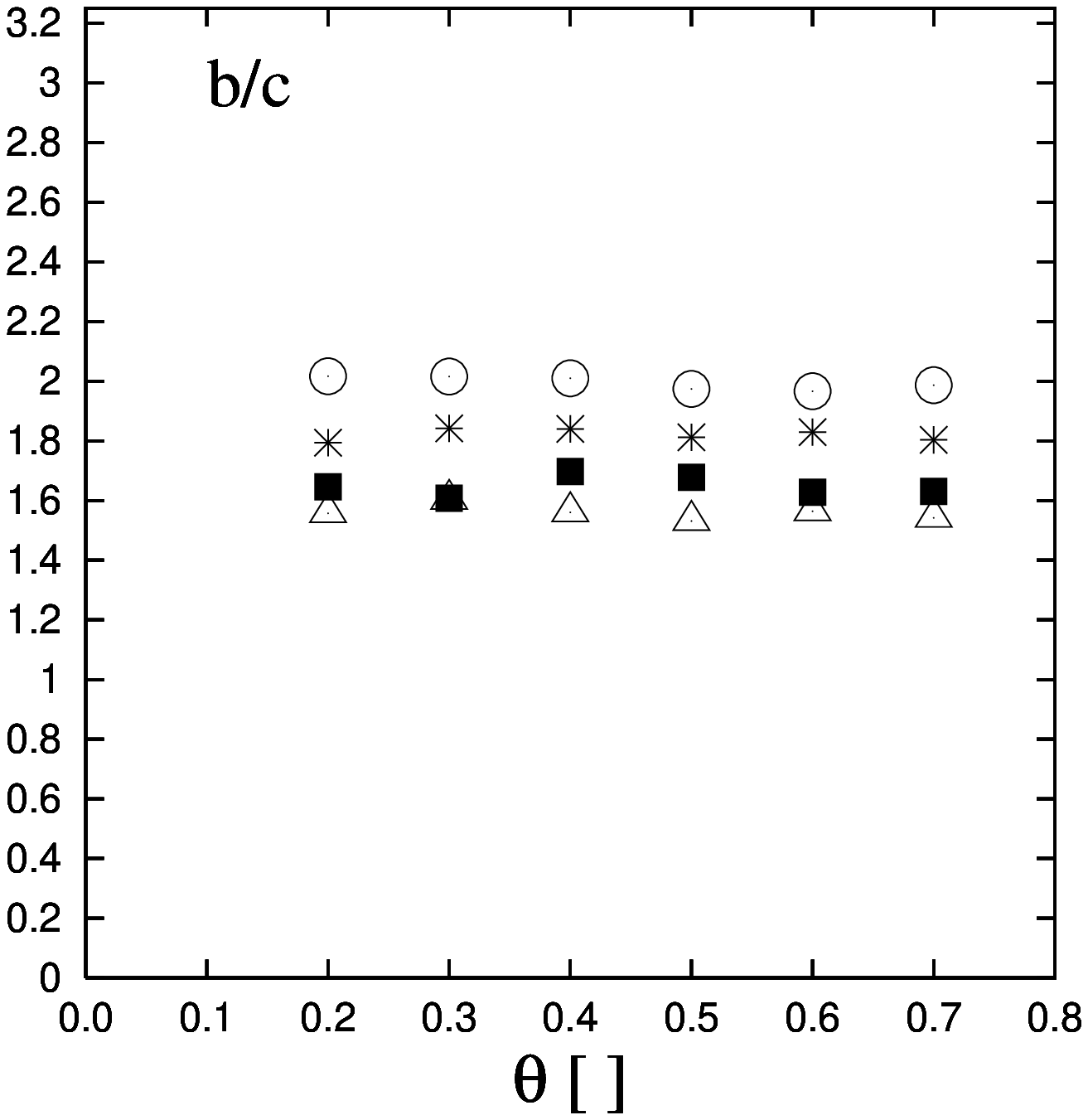} 
}
	\put(1.5,1.5){ \epsfysize=2.5in
		    \epsffile[ 100 150 450 600]
		   {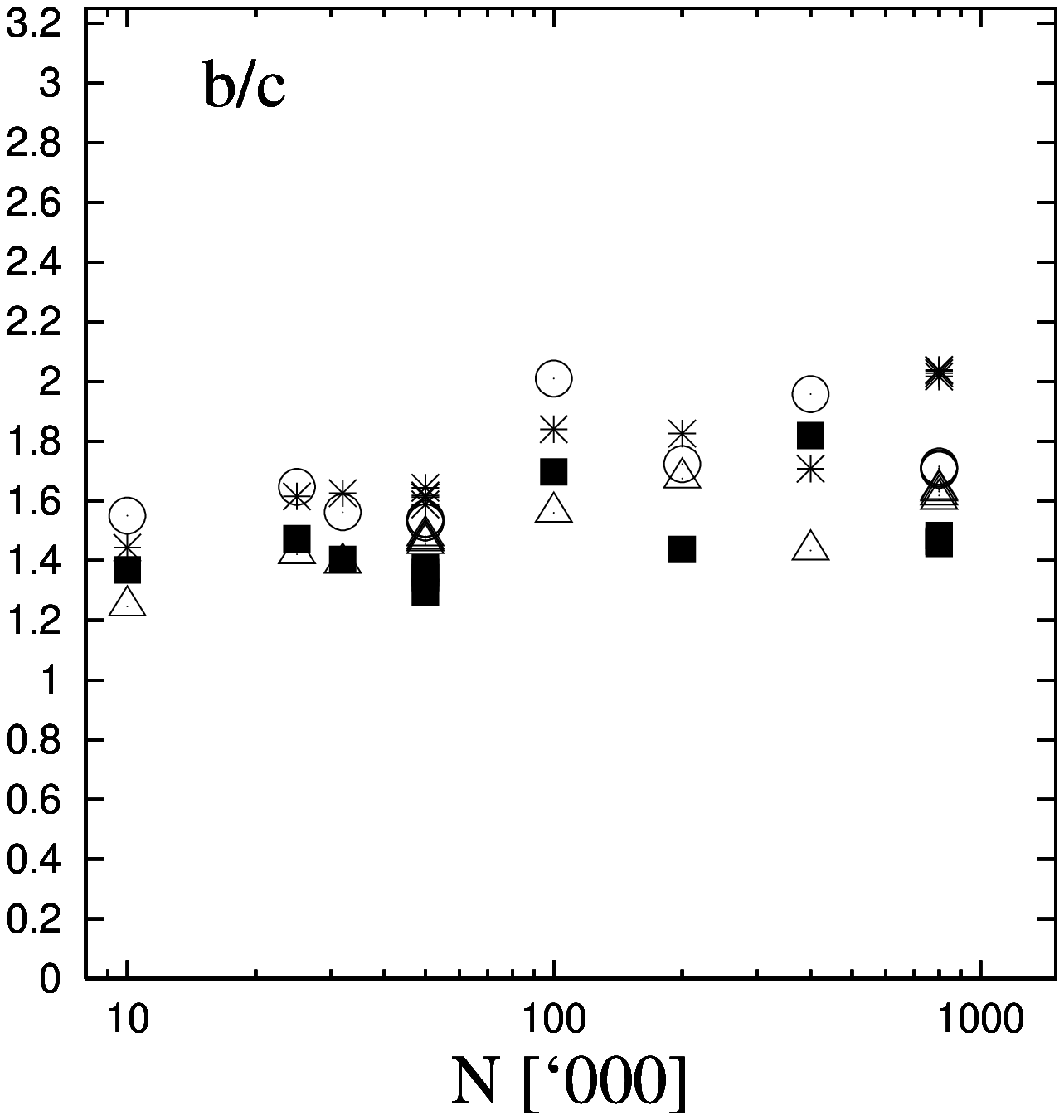} 
                  }	
	\put(1.5,7.25){ \epsfysize=2.5in
		    \epsffile[ 100 150 450 600]
		   {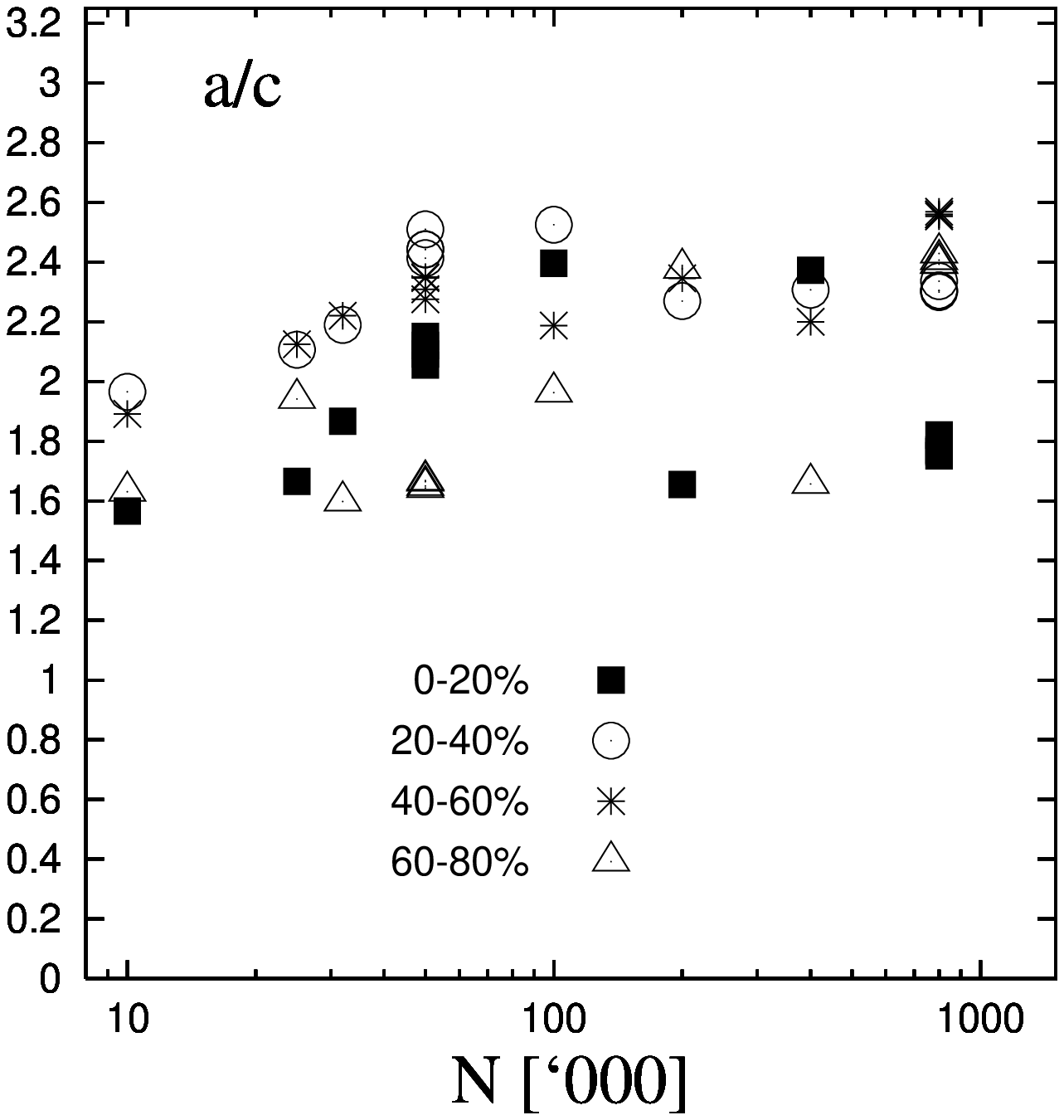} 
		}
\end{picture} 
      \caption{Axis ratios  $a/c$ and $b/c$  as 
function of the  opening angle $\theta$ (right-hand panels), particle number  $N$ 
(middle panels) and  resolution $1/\epsilon$  (left-hand panels). 
Simulation parameters are listed in Tables~\ref{tab:N} and \ref{tab:Epsilon}. 
Values obtained for each of four 20\% mass bins are displayed with different symbols.} 
         \label{fig:AC}
   \end{figure*}


   \begin{figure*}
\setlength{\unitlength}{1cm} 
\begin{picture}(8,8)(0,0) 

	\put(-3.5,1.75){ \epsfysize=0.4\textwidth
		    \epsffile[ 100 150 450 600]
		   {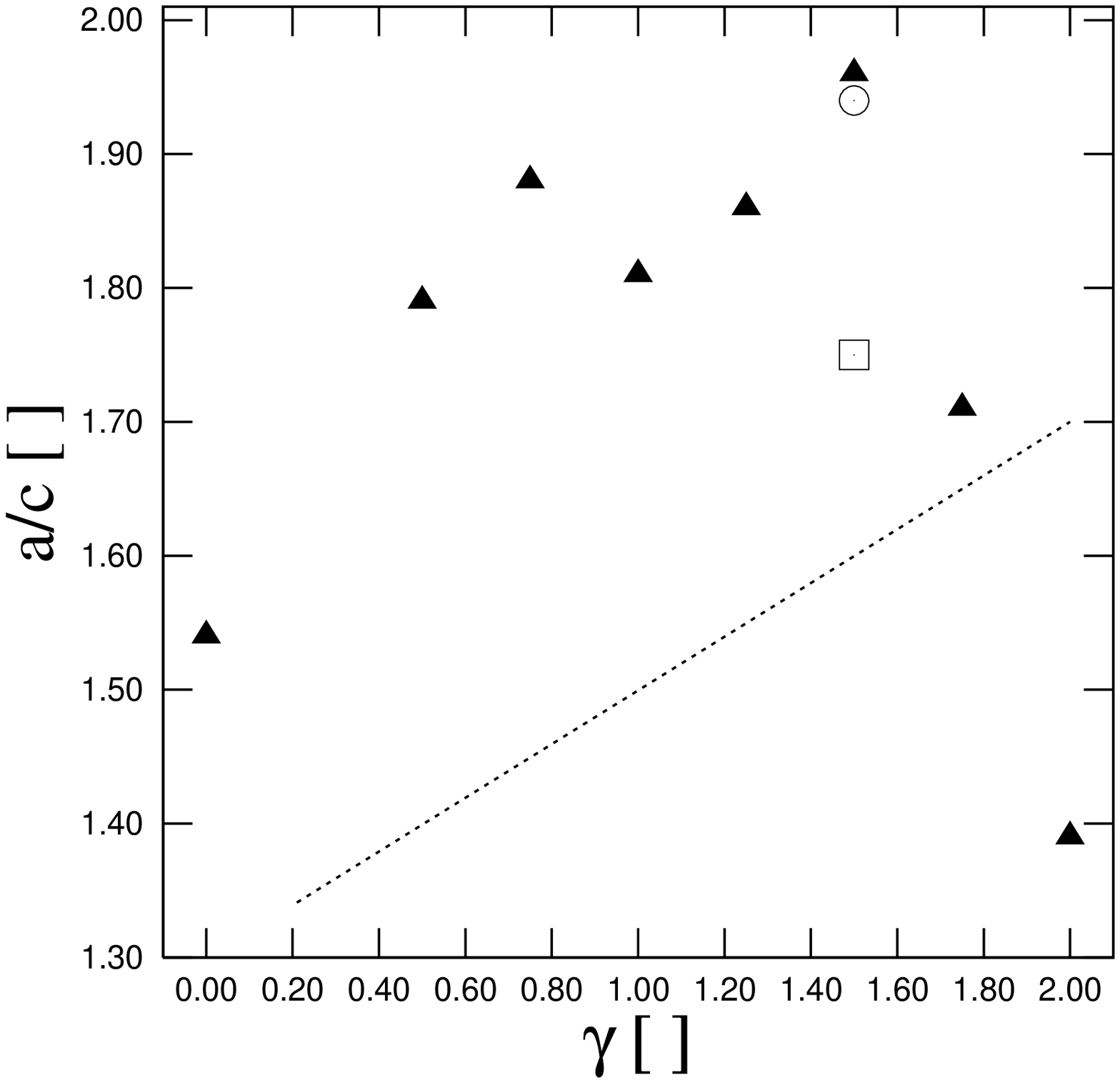}  
}
	\put(5.,1.75){ \epsfysize=0.41\textwidth 
		    \epsffile[ 100 150 450 600]
		   {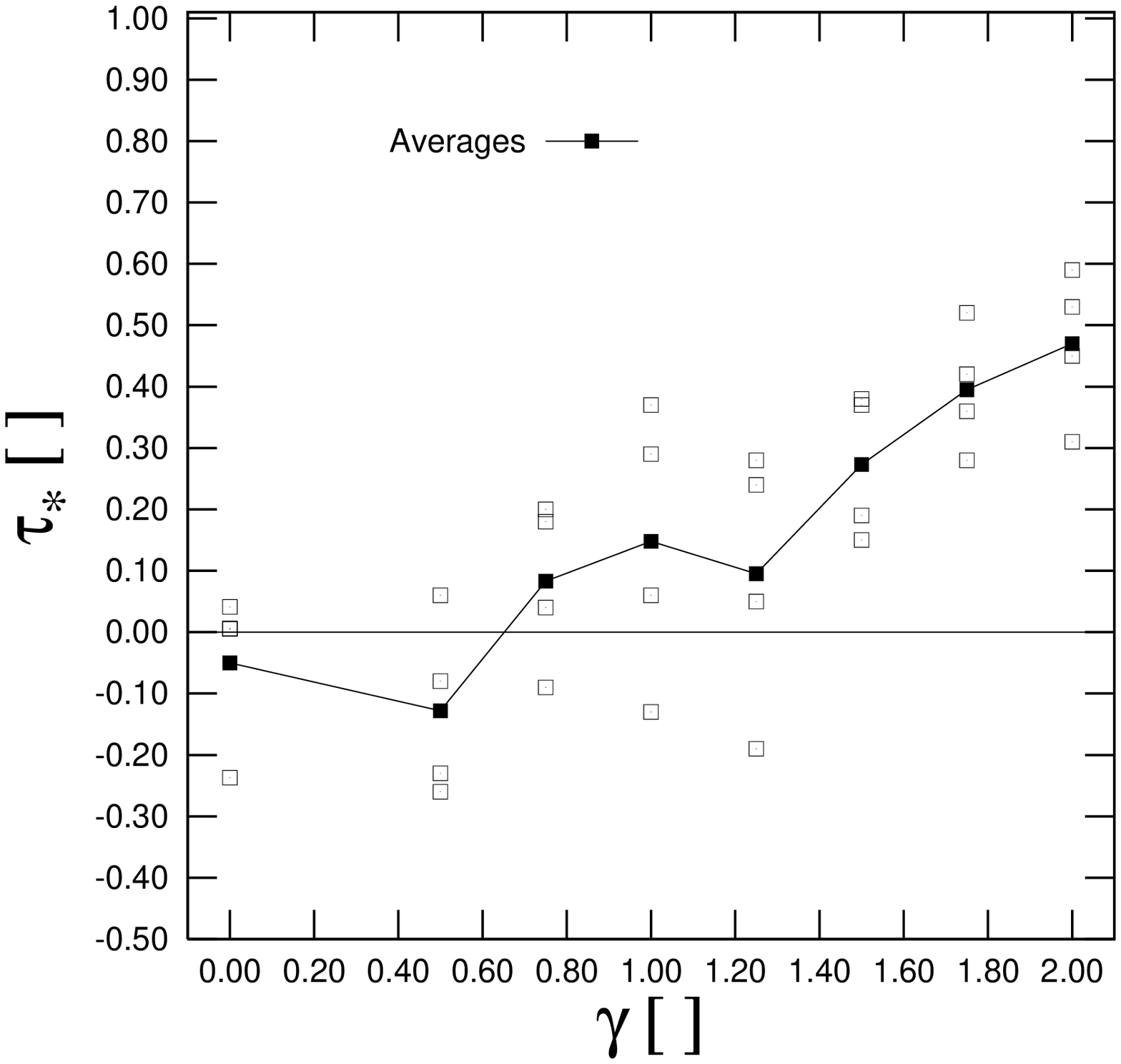} 
                  }
		\put(8.5,2){ {\small} {\Huge $\star$} } 
		\put(10.1,1.8){ {\Huge $\dagger$} } 
\end{picture} 
      \caption{(A, left-hand panel) Ratio of major axis $a$ to shortest axis $c$ for runs with $N = 100,000$ particles ($\blacktriangle$)  
and different power index $\gamma$. Results with 10,000 ($\Box$) and 25,000 $(\bigcirc$) particles are also displayed for the 
case $\gamma = 3/2$.  The straight dotted line is the fit $a/c = \gamma/5 + 1.3$ lifted from C\&H+92. 
\label{fig:acvsgamma}
(B, right-hand panel) Shape parameter $\tau_\ast$ versus initial power index $\gamma$. 
The horizontal  line indicates the boundary between oblate (positive) and prolate shapes.
On the mean the structures are mostly oblate, more so for $\gamma > 1$.  Averaged values for each 
$\gamma$ are linked with solid lines. The star marks the average result of 5,000 particle $\gamma = 1$ runs from A\&M+90; the dagger at $\gamma = 3/2$ marks the average result of 10,000-particle runs by C\&H+92. }
         \label{fig:Tauvsgamma}
   \end{figure*}

\subsection{Smoothing length issues} 
\label{subsec:softening}

If $\epsilon$ is large we expect a loss of resolution and a bias in the 
outcome of the calculations.  Burkert (1990)  and Boily et al. (1999) have discussed this in terms of 
a cut off in orbit deflection angle from the mean field and artificial saturation of the phase-space density. We looked for such biases in a series of calculations with
varying smoothing length. 

Figures~\ref{fig:Tau} and \ref{fig:AC} (left-hand panels) graph the axis ratios and the 
averaged values of  $\tau_\ast$  for the 
runs listed in Table~\ref{tab:Epsilon}. All runs had $N = 100,000$ particles. 
The averaging was done for the 80\% most bound particles to 
 facilitate comparisons with C\&H+92. The  error bar displayed  (of 0.22, bottom-left panel) 
is scatter. The trend seen  in the data clearly  indicates that 
 higher resolution runs yield more oblate structures.  Note the apparent convergence 
achieved for $1/\epsilon = 256$, as $\tau_\ast$ and the axis ratios  flatten out around that value. 

The results of Fig.~\ref{fig:Tau} and \ref{fig:AC} underline the importance of using both  sufficient number of particles and  resolution for convergence of the equilibrium morphology. By contrast, 
the equilibria are not sensitive to the value chosen for
the  opening angle $\theta$ (Fig.~\ref{fig:AC}, right-most 
panels). If we compare values of $\tau_\ast$ at 
$1/\epsilon = 10$ with those for $1/\epsilon = 512$, we find a systematic increase of $\approx + 0.3$, enough to bridge the gap 
between our results  for $N = 10,000$ particles and those of C\&H+92 
seen on the right-hand  panel of Fig.~4.  Thus, the reduced resolution of their calculation 
accounts for the most part for the strong prolate morphology of their equilibria. 

\subsection{Morphology in relation to $\gamma$} 
The sensitivity of equilibrium parameters to numerical resolution  suggested to us to seek 
 a relation between equilibrium morphology and initial density profile with higher-resolution 
 calculations than was done in earlier work. 
On Fig.~\ref{fig:acvsgamma}a we graph the ratio of major- to shortest-axis $a/c$ as function of 
the power index $\gamma$. Table~\ref{tab:Gamma} lists the parameters of the $N = 100,000$ runs. 
 The graph shows no trend with $\gamma$, with values of $a/c$ (here  
averaged over the 80\% most-bound particles) ranging between l.4 and 2. On the figure we added the 
linear fit to the data from C\&H+92 for the same quantity for comparison. The results of an 
$N = 10,000$ and an $N = 25,000$ calculation, with the same linear resolution 
$\epsilon = 1/512$, are also added in Fig.~\ref{fig:acvsgamma}a. The 10,000-particle result falls close to the straight line from C\&H+92,  
while the 25,000-particle run gives a ratio comparable to those obtained with 
100,000-particle calculations. This confirms our expectation that 10,000-particle calculations 
do not quite reach maximum potential during violent relaxation (cf. BAK+02) 
and as a result do not achieve in equilibrium the  same final (converged) morphology
of larger-$N$ calculations. 
Note that the axis ratios for the cases $\gamma = 0$ and  $\gamma = 2$ differ significantly from those obtained with other power indices. 
Nevertheless, the full range displayed on the y-axis of Fig.~\ref{fig:acvsgamma}a 
brackets E2-E5 morphological types, which are by no means exceptional values 
 when compared with those of elliptical galaxies. 
 
 On Fig.~\ref{fig:Tauvsgamma}b we plot the values of $\tau_\ast$ as a function 
 of initial power index $\gamma$. As before, the systems 
 were split in equal 20\%-mass shells to compute $\tau_\ast$ at times ranging from 60 to 80 $N$-body units. The 
 scatter seen on Fig.~\ref{fig:Tauvsgamma}b results form  fluctuations between 
 different mass bins. The solid line and filled symbols indicate averages.  
 In contrast to the results for the aspect ratios,  
 we find a well-defined trend of the parameter $\tau_\ast$ versus power index $\gamma$ (Fig.~\ref{fig:Tauvsgamma}b) such that $\tau_\ast$  increases 
 on the mean with  increasing $\gamma$, and $\tau_\ast > 0$ 
 for $\gamma \ge 5/4$. The case with 
$\gamma = 2$ yields the most oblate and  axisymmetric equilibrium of all the cases we have explored. 
 Data points for $\gamma = 1$ and lower are only mildly oblate on the mean, while the case $\gamma = 1/2$ in fact gave a mildly prolate structure.  
 However,  the morphology of all runs with $\gamma \ltabout 3/4$ is 
 consistent with axisymmetry  or perfect-ellipsoid morphology 
 ($\tau_\ast = 0$), owing to the scatter. Seen in this light, we would conclude 
 that the morphology of systems with $\gamma \ge 3/2$ is oblate for all mass
 shells, whereas those with lower $\gamma$ values are a mixture of oblate and prolate shells. 
 

   \begin{figure*}
\setlength{\unitlength}{1cm} 
\begin{picture}(8,8)(0,0) 
	\put(-4.,2.){ \epsfysize=0.4\textwidth 
		    \epsffile[ 100 150 450 600]
		   {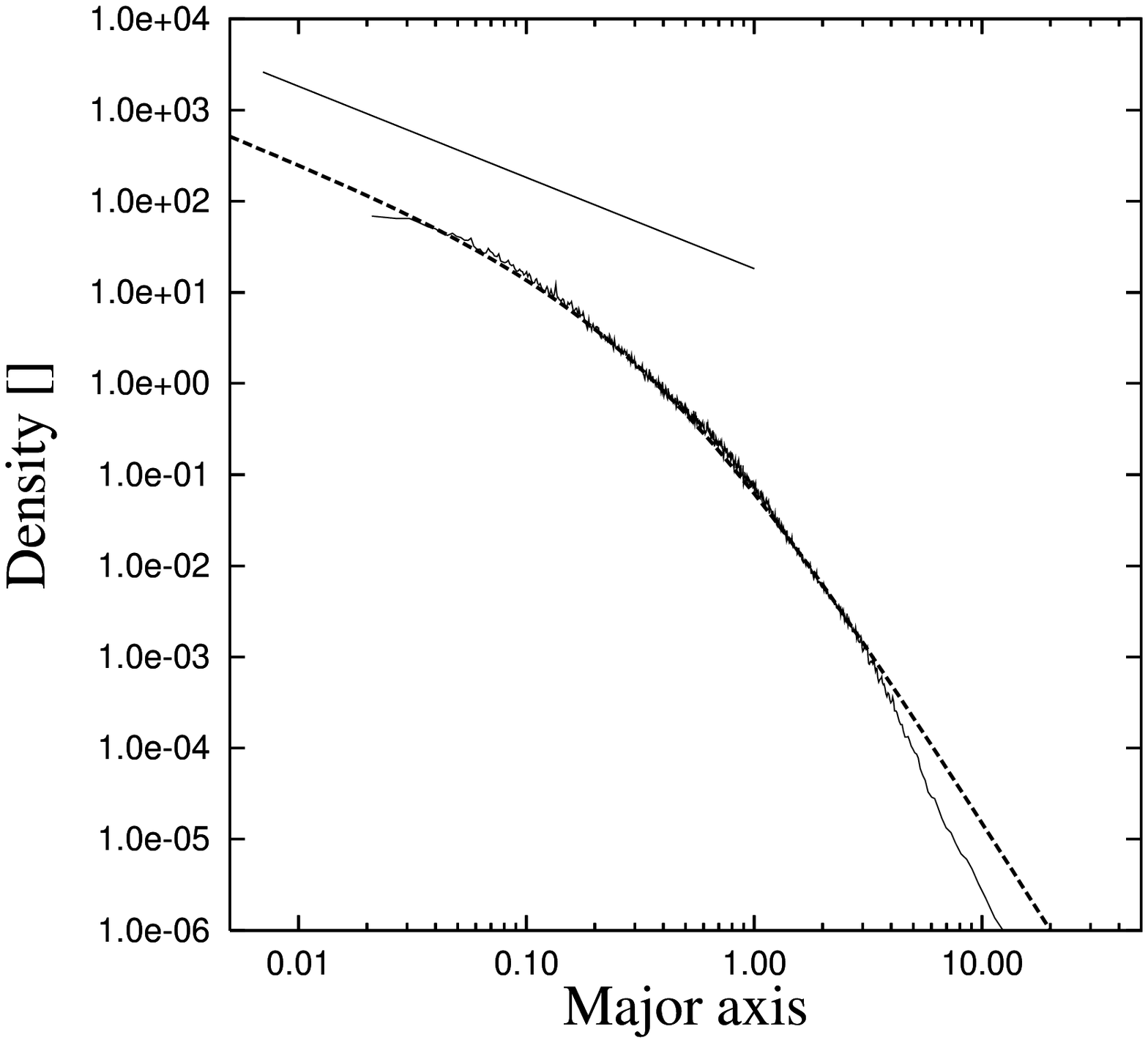} }
 	\put(-.3,2){ \Large $r_h$ }  \thicklines \drawline[+40](-.0,1.7)(-.0,1.2)  \drawline[+40](-.0,2.5)(-.0,4.25) 

	\put(5.05,2.5){ \epsfysize=0.6\textwidth 
		    \epsffile[ 100 150 450 600]
		   {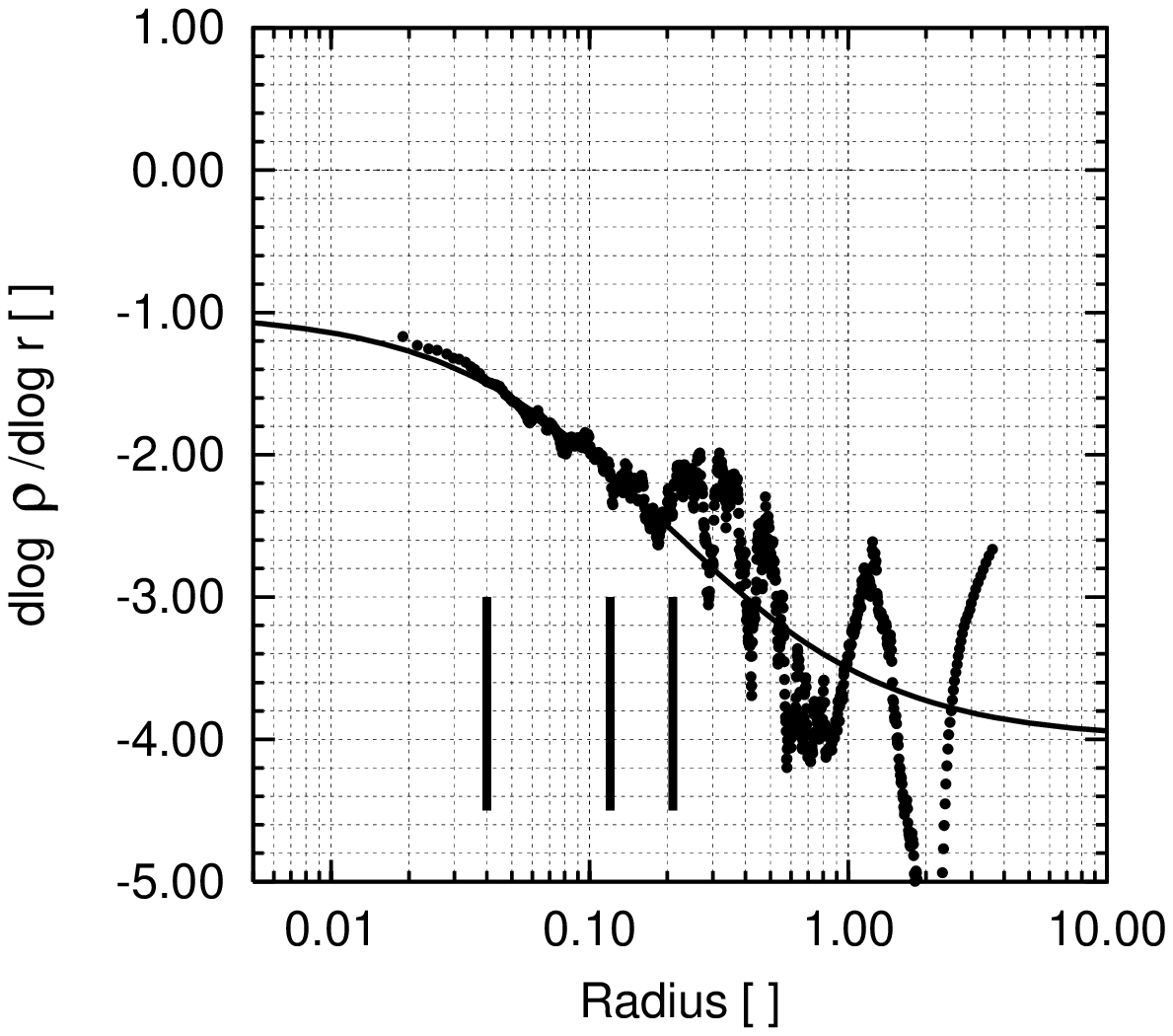} 
		}
\end{picture} 
      \caption{(A, left-hand panel) The equilibrium density profile of run d013 versus semi-major axis length. The dashed line is a fit of 
the data with an Hernquist model (Eq.~\ref{eq:Hernquist}). 
 The straight line in the upper-left has a slope of -1 and the half mass radius $r_h$ is given by a vertical tick mark. 
         \label{fig:Fithernquist}
(B, right-hand panel) 
The logarithmic derivative of the density profile shown on panel (A). 
 The solid line is obtained from Eq.~\ref{eq:Hernquist}.  The vertical 
tick marks indicate (from left to right) the 1\%, 10\% and 20\% most bound mass fraction.  \hspace{2.75cm} $\, $  } 
         \label{fig:dlogHernfit}

   \end{figure*}


   \begin{figure*}
\setlength{\unitlength}{2cm} 
\begin{picture}(6,11)(0,0) 
	\put(-.75,8.45){ \epsfysize=4.2in
		    \epsffile[ 100 150 450 600]
		   {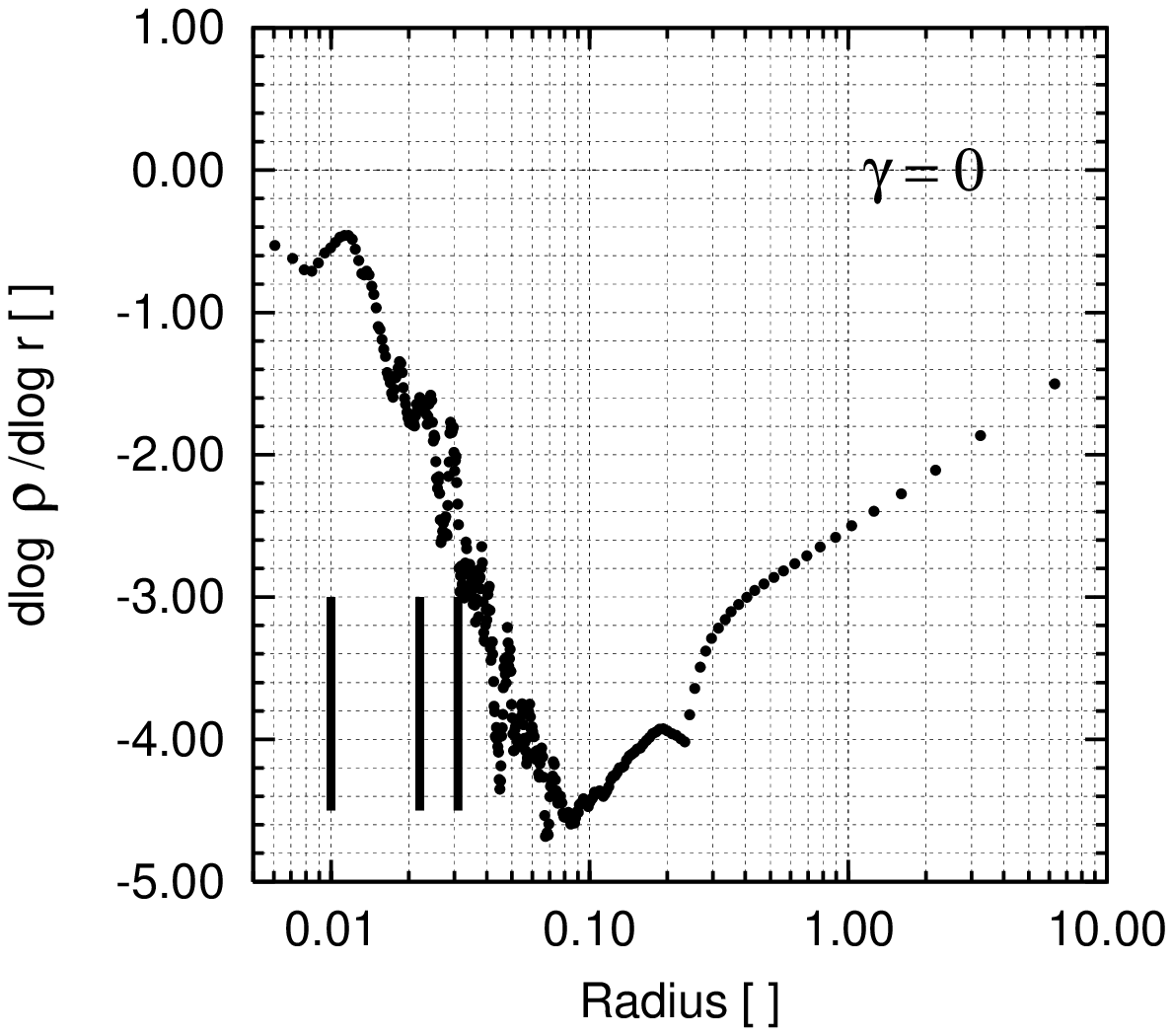} 
}
	\put(-.75,4.85){ \epsfysize=4.2in
		    \epsffile[ 100 150 450 600]
		   {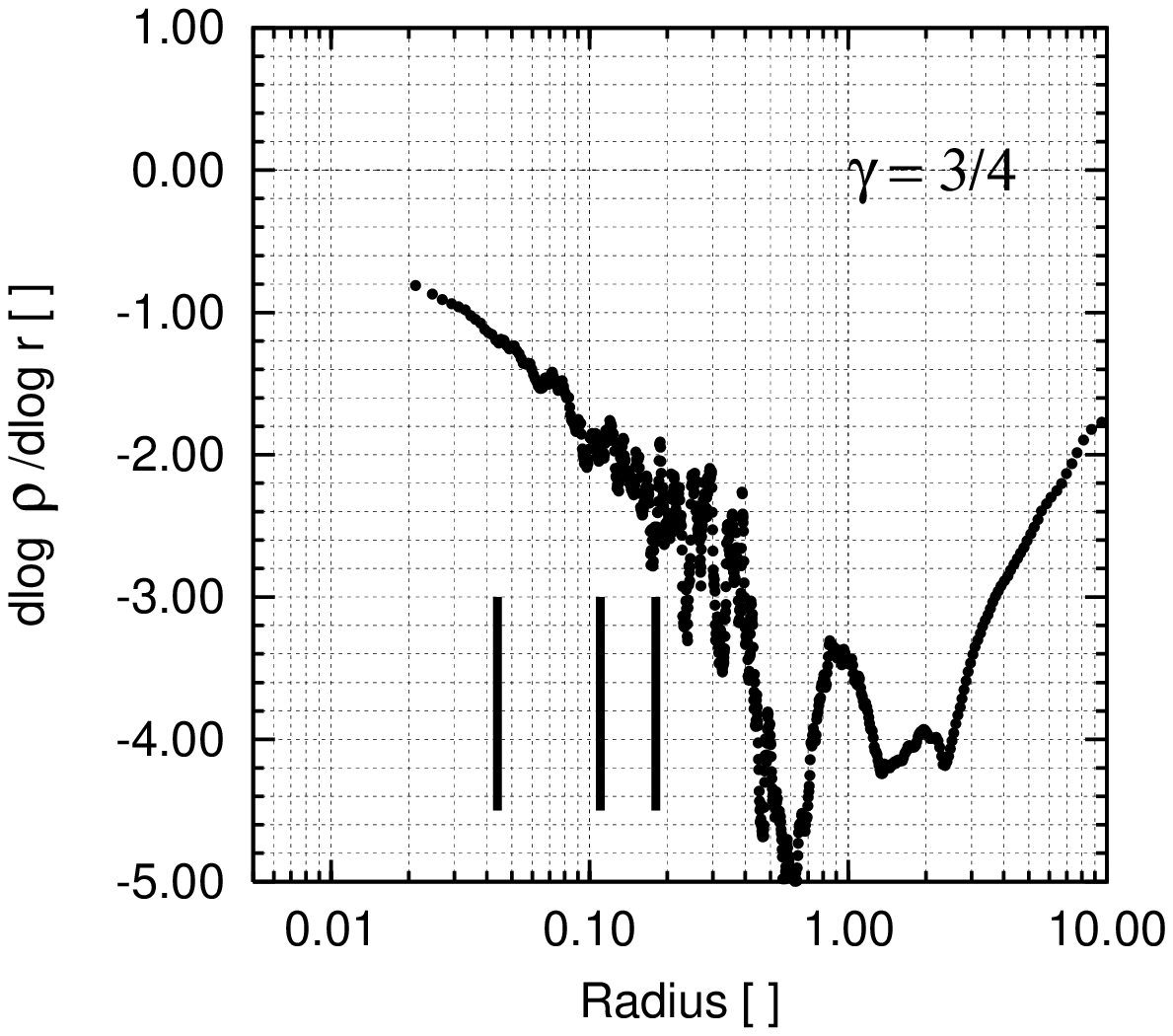} 
                  }	

	\put(-.75,1.25){ \epsfysize=4.20in
		    \epsffile[ 100 150 450 600]
		   {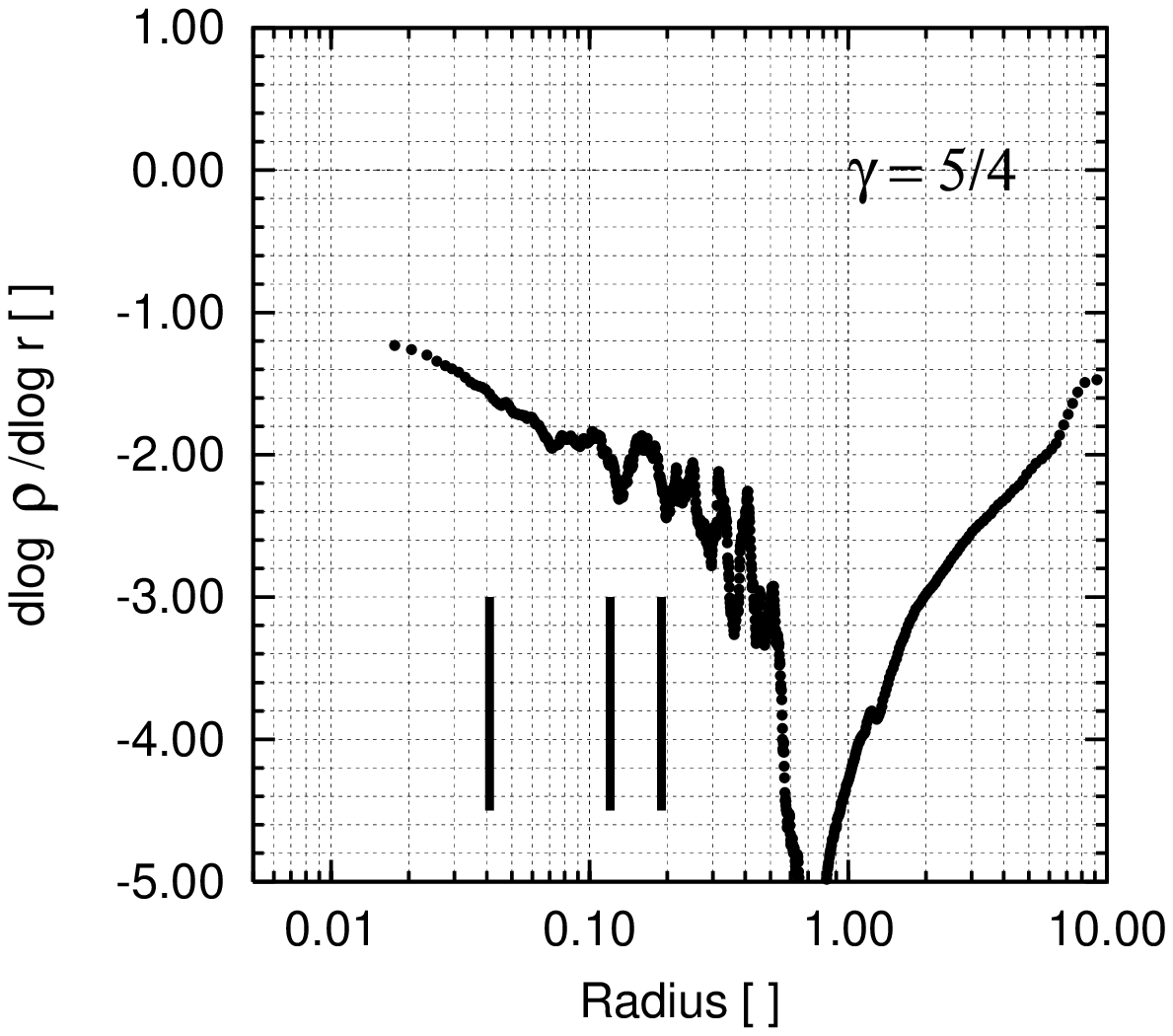} 
			}
	\put(3.6,1.25){ \epsfysize=4.20in
		    \epsffile[ 100 150 450 600]
		   {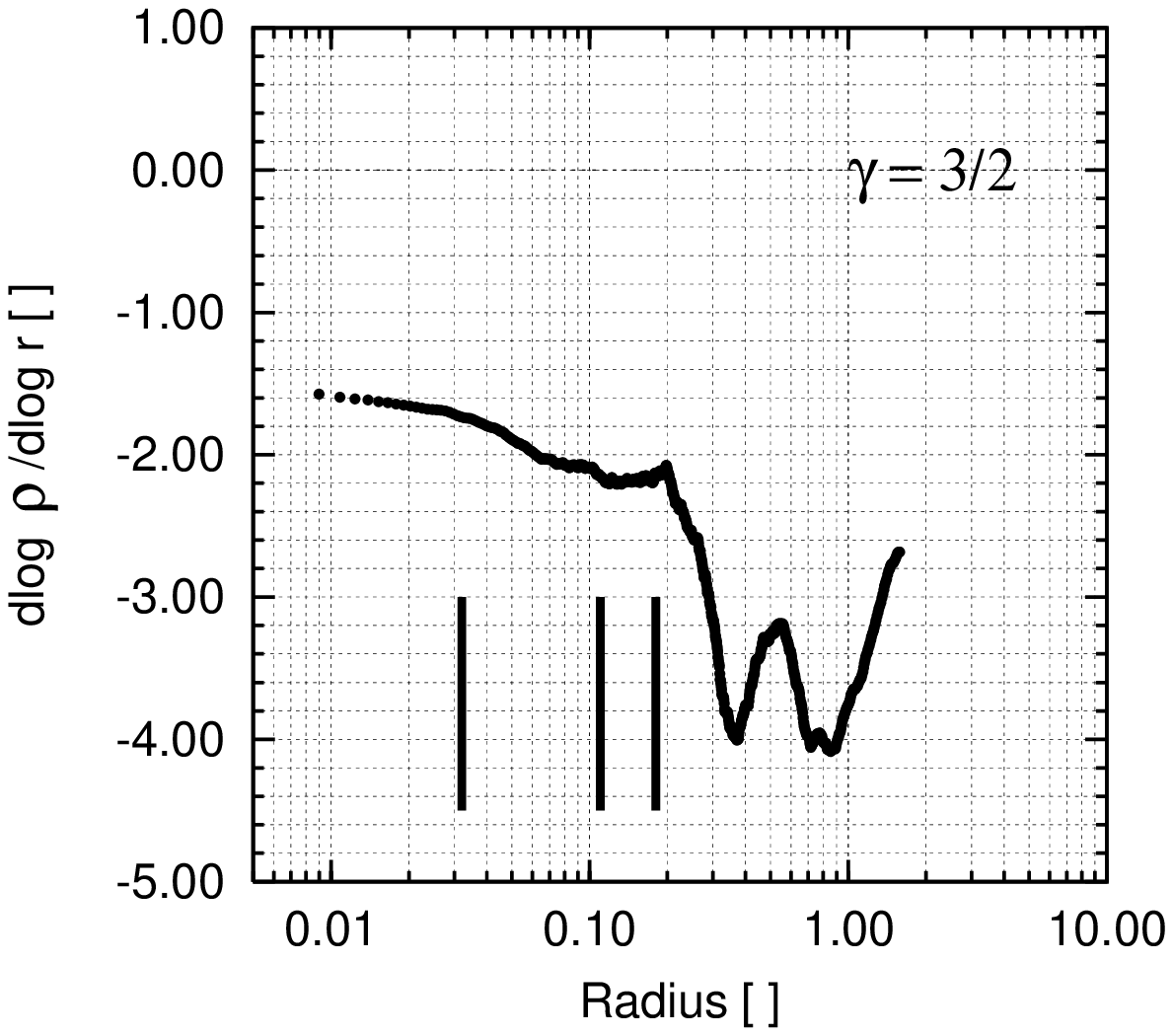} 
}
	\put(3.6,4.85){ \epsfysize=4.20in
		    \epsffile[ 100 150 450 600]
		   {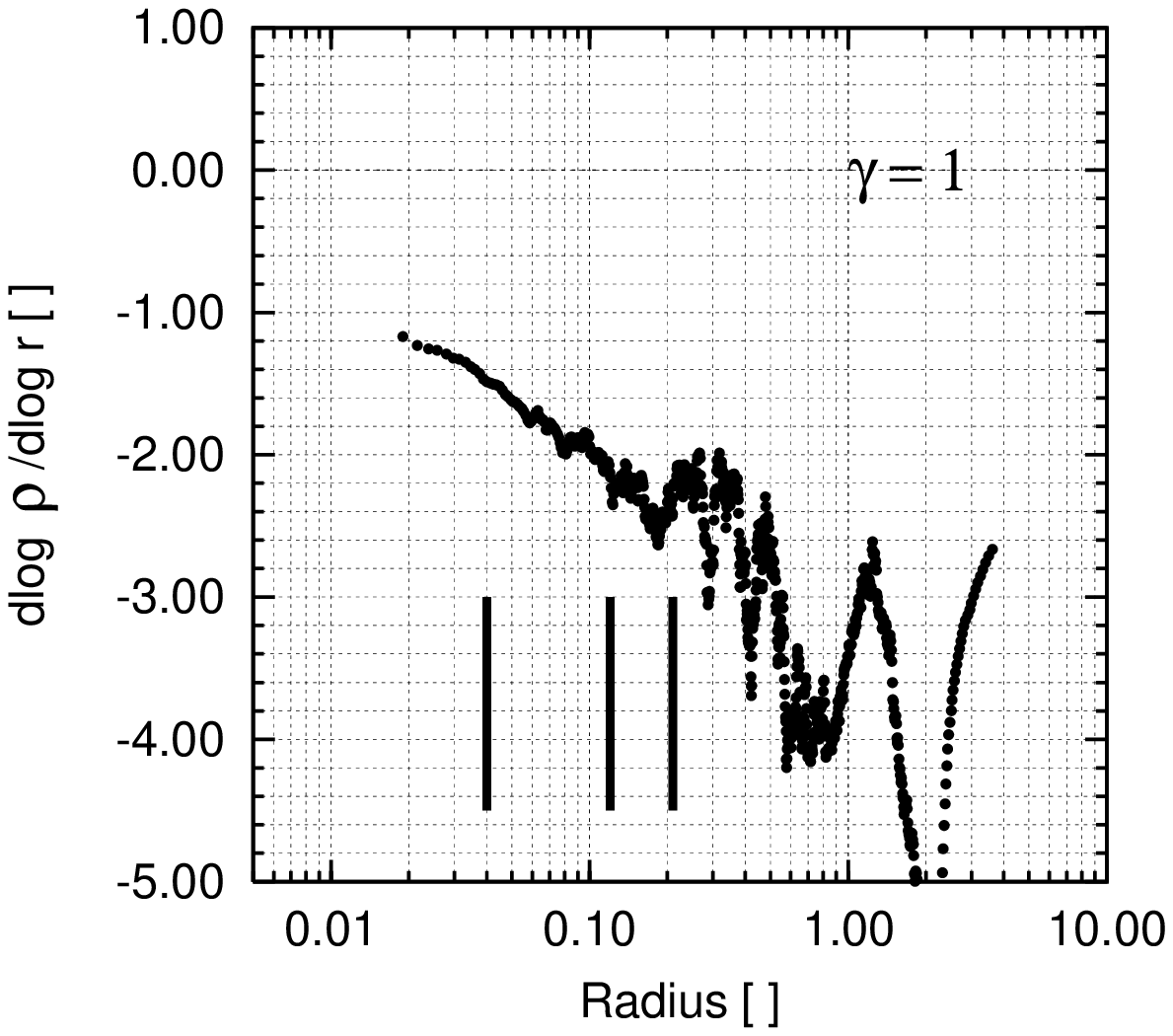} 
                  }	
	\put(3.6,8.45){ \epsfysize=4.20in
		    \epsffile[ 100 150 450 600]
		   {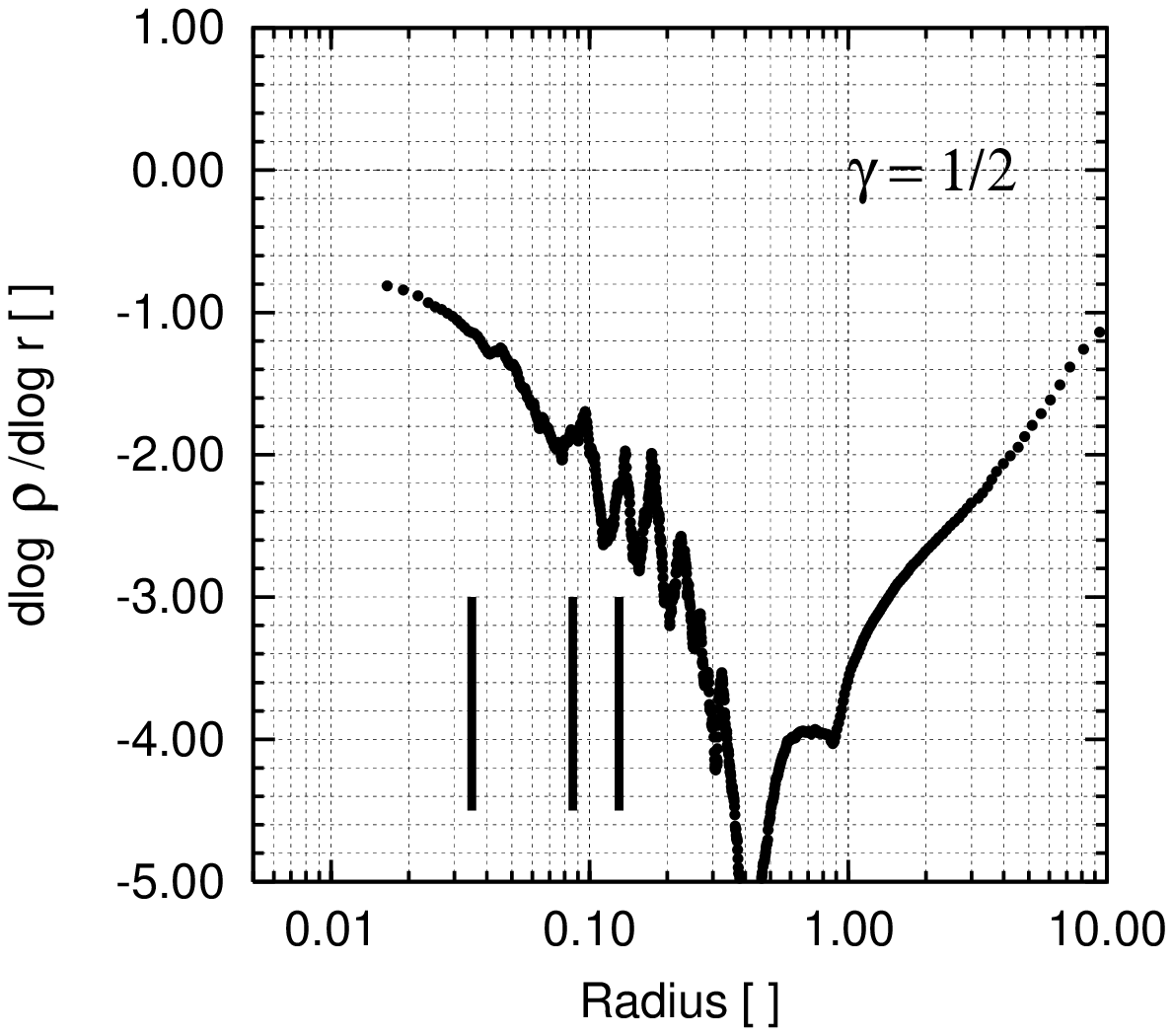} 
}
\end{picture} 
      \caption{The logarithmic derivatives for equilibria obtained from different initial power-law index $\gamma$. $N = 100,000$ in all cases but for the case $\gamma = 3/2$ which had $N = 800,000$ particles. The vertical 
tick marks indicate (from left to right) the 1\%, 10\% and 20\% most      bound mass fraction.} 
         \label{fig:DlogRho}
   \end{figure*}

\setcounter{figure}{7} 

   \begin{figure*}
\setlength{\unitlength}{2cm} 
\begin{picture}(6,4)(0,0) 
	\put(-.75,1.25){ \epsfysize=4.20in
		    \epsffile[ 100 150 450 600]
		   {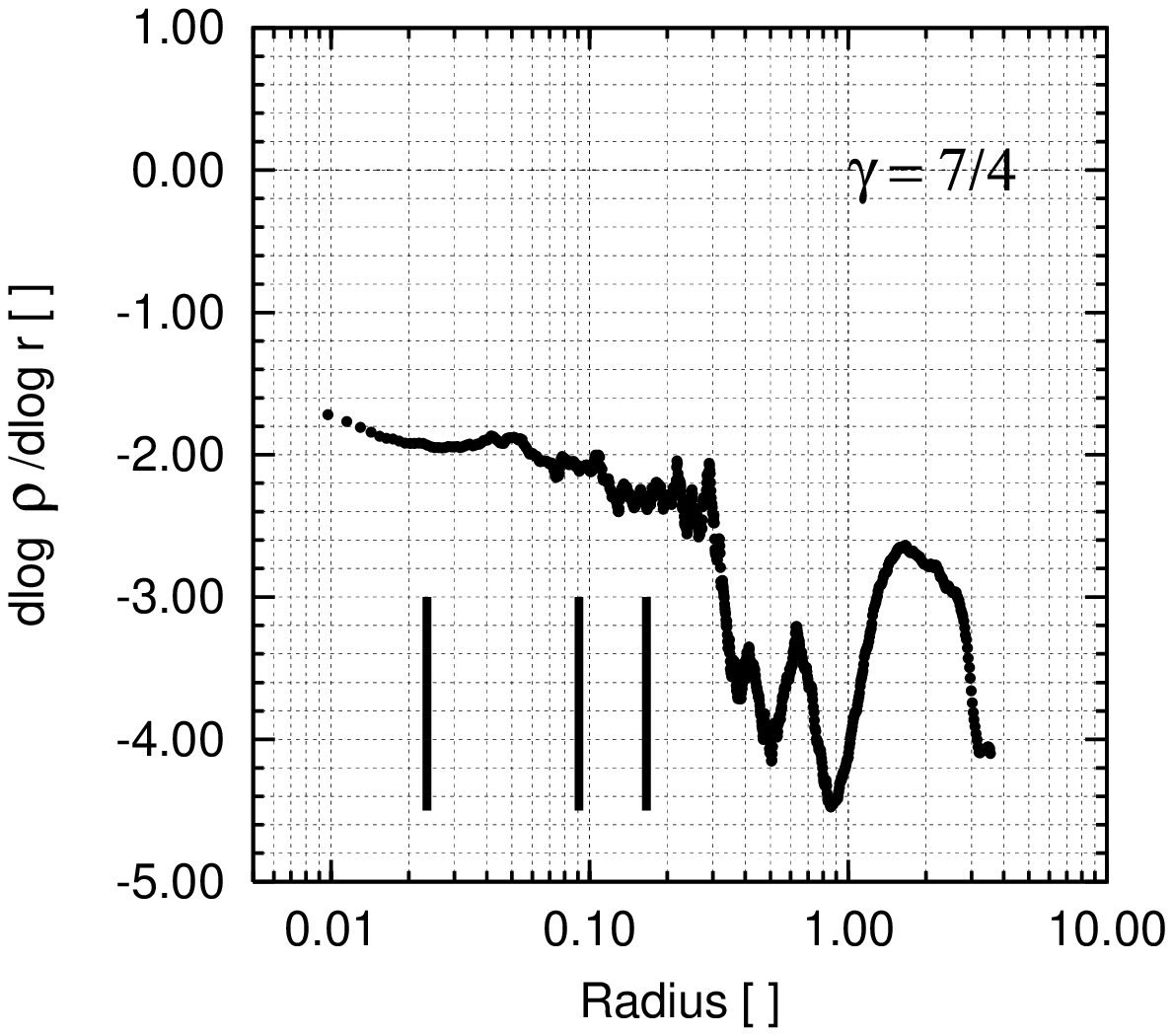} 
			}
	\put(3.6,1.25){ \epsfysize=4.20in
		    \epsffile[ 100 150 450 600]
		   {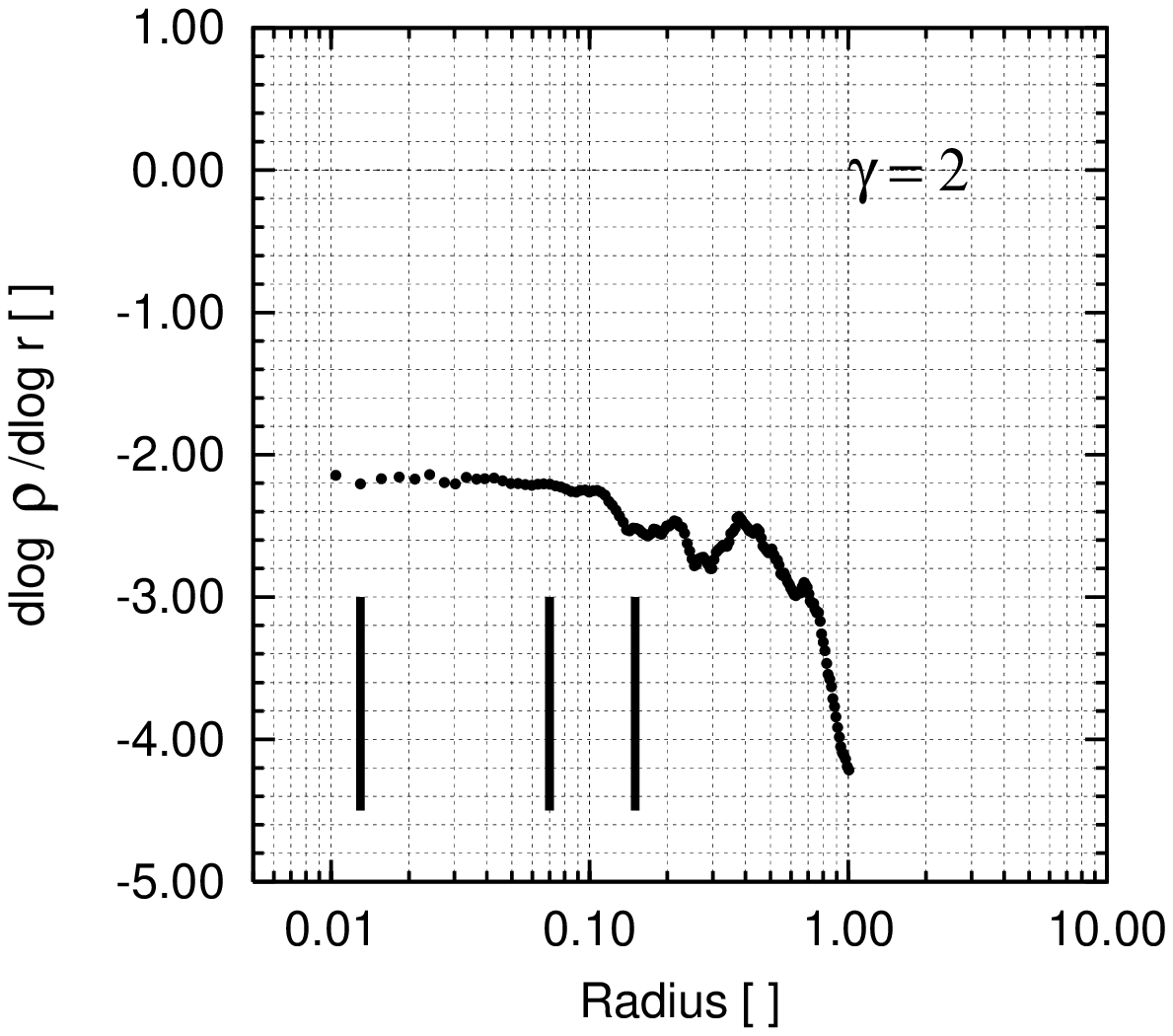} 
}
\end{picture} 
      \caption{{\it -- continued} } 
   \end{figure*}



   \begin{figure*}
\setlength{\unitlength}{1cm} 
\begin{picture}(12,14)(0,0) 
	\put(-1.,2.){ \epsfysize=.5\textwidth
		    \epsffile[ 100 150 450 600]
		   {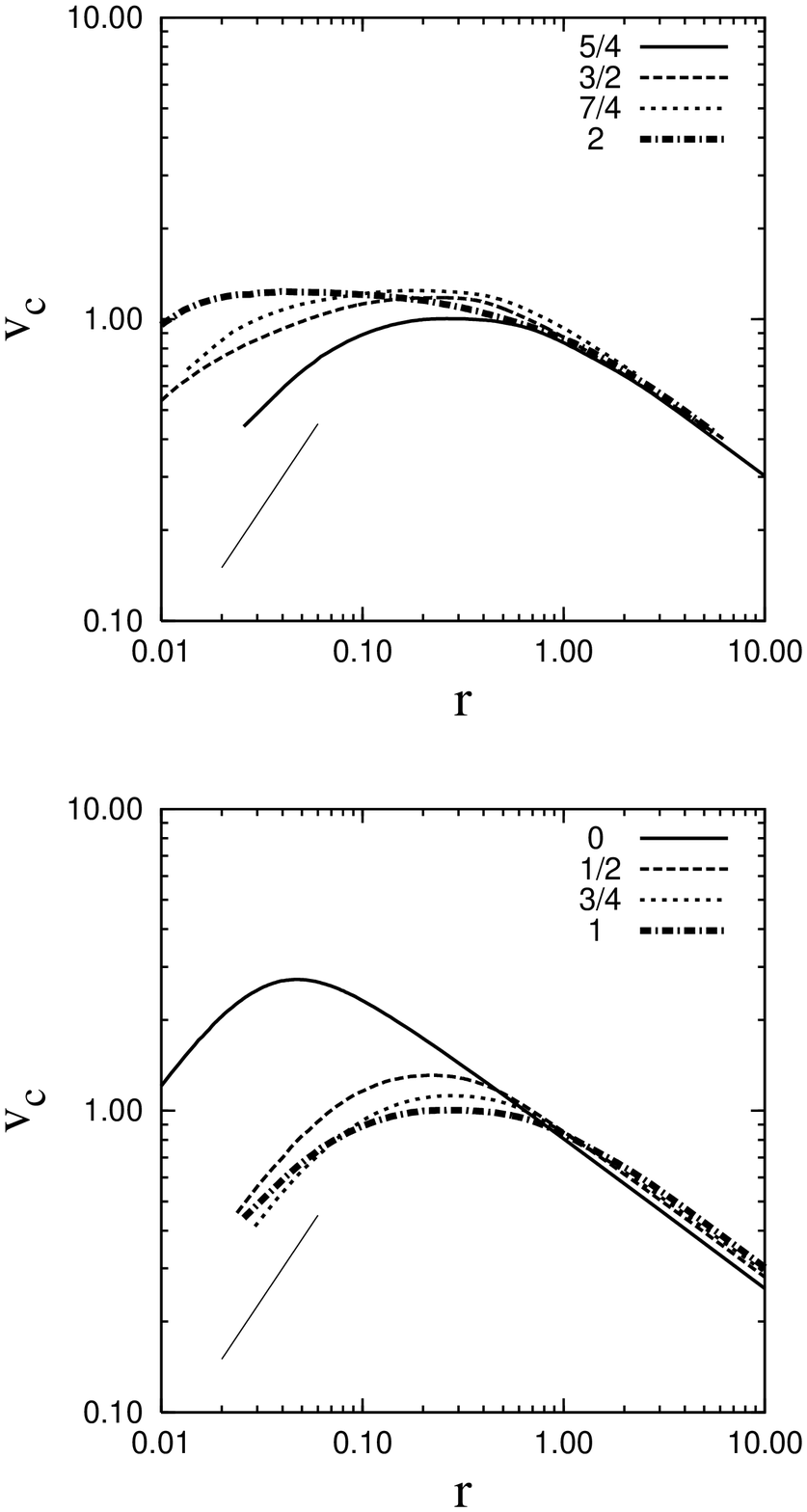} 
			}
	\put(6.7,2.){ \epsfysize=.5\textwidth
		    \epsffile[ 100 150 450 600]
		   {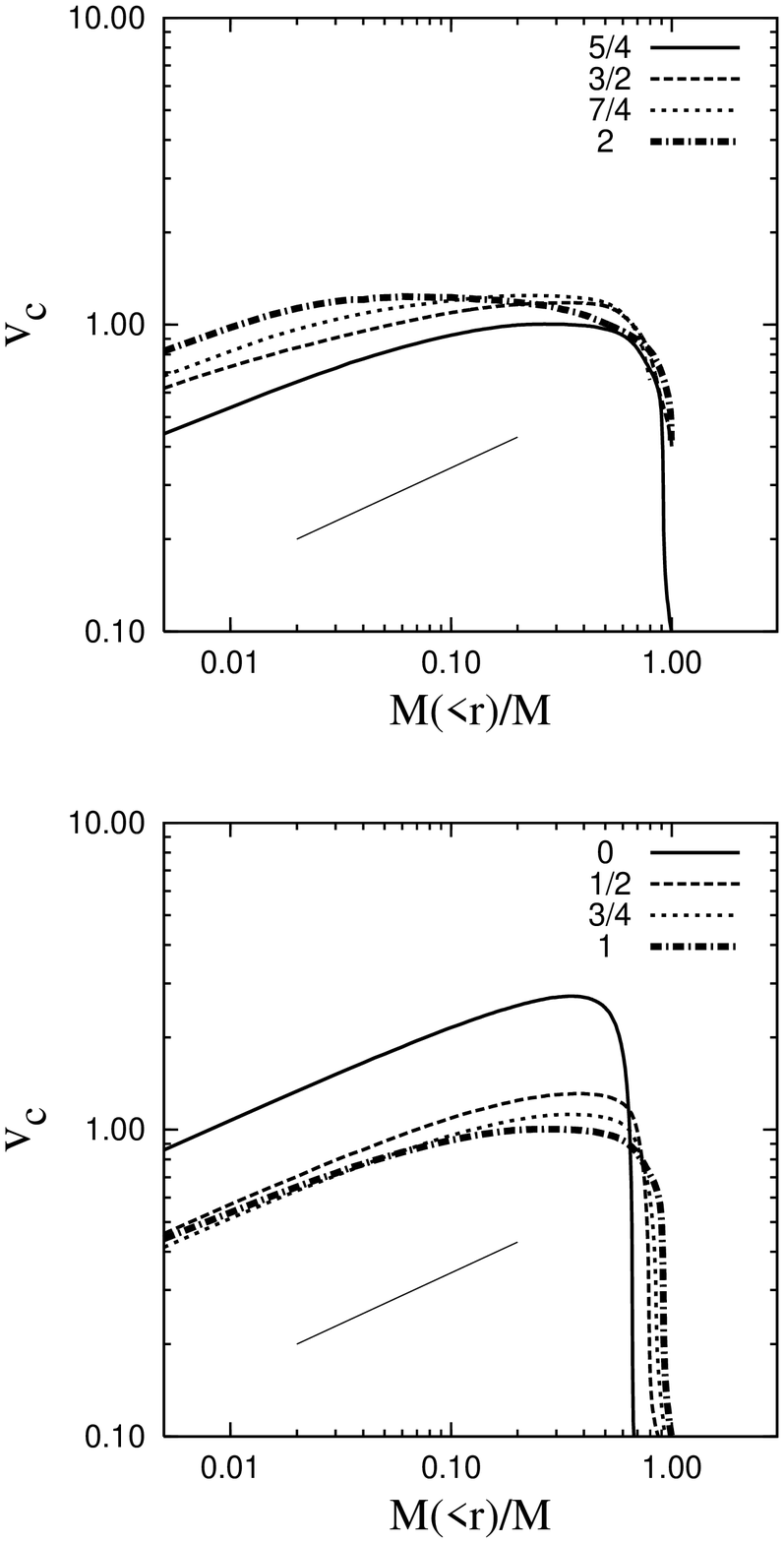} 
}
\end{picture} 
      \caption{Rotational velocity $v_c$ plotted for different $\gamma$ as function of radius (left-hand panels) and 
of integrated mass $M(<r)$ (right-hand panels). Lower set : $0 < \gamma \le 1$. Higher set : $ 1 < \gamma \le 2$. The straight lines 
are $\propto r$ (left-hand panels) or $\propto M^{1/3}$ (right-hand panels) corresponding to a 
constant mass density.}
         \label{fig:vc}
   \end{figure*}

\section{Global and inner density profiles} 
Fig.~\ref{fig:Rhovsr} displays typical spherically-averaged density profiles for our simulations with $\gamma = 3/2$. 
Our basic expectation from incomplete relaxation 
is that the density profile will match the power-law of the initial conditions on small scales. 
Power-law asymptotes at small and large radii 
 are a generic feature of dissipation-less relaxation in cosmology (Navarro et al. 1997; Moore et al. 1998). It is 
natural to expect them also in the present, more restraint, context. We therefore set out to fit the radial density profiles 
of our equilibria with a continuous function of radius. All density profiles where averaged over spherical 
shells containing 100 particles. We varied the number of particles per shell (200, 500) but this did not lead to significant improvements, save for the very-large $N$ d007 
model where $500$ particles still allowed to probe 
a sufficient range in radius while keeping the noise level low. 

 C\&H+92 had found a good fit to their equilibrium  
$\gamma = 1$ run averaged over spherical shells using  an Hernquist (1990) density distribution, 

\begin{equation} \rho(r) = \frac{M}{2\pi}\frac{r_0}{r}\frac{1}{(r_0+r)^3}, \label{eq:Hernquist}\end{equation} 
where $M$ is the total system mass, and $r_0$ a free length. 
We also fitted the equilibrium profile obtained from an 100,000-particle $\gamma = 1$ run  (d013, 
cf. Table~\ref{tab:IC})  with (\ref{eq:Hernquist}). 
We first evaluated the density locally following the scheme of Casertano \& Hut (1985) as coded in the NEMO 
analysis package. The rms axis ratios  were found as in \S3, again covering the innermost 80,000 particles only. 
This allowed us to 
  round up the distribution, by shifting the particles  in x-y-z {\em inversely} as 
the axial ratios obtained~: thus  the triaxial configuration was straightforwardly transformed to a sphere of the same volume. 

The result is shown on Fig.~\ref{fig:Fithernquist}a. 
The curve shown has  $r_0 = 0.288$ with $M = 1$. 
Note that the half-mass radius ($r_h \simeq 0.53$) obtained numerically is significantly less than 
the Hernquist function $= (1+\sqrt{2})\,r_0 = 0.69(5)$ owing to the mass deficit at large distance.  Inside $r \simeq 10$  the mass integrated 
from (\ref{eq:Hernquist}) is 92.5\% of the total, while the mass found numerically  reaches 91.2\% of 
the total. However , 
the data and analytic curve are in complete disagreement for $r > 10$. 
Nevertheless a comparison of this fit to the one 
 shown on  Fig.~8 of C\&H+92 
 indicates that our larger-$N$ simulations give an ever closer agreement to the
Hernquist profile. The fact that both profiles are relatively well fitted by the
Hernquist law is unexpected, since the morphology
 of the two equilibria are very different from each other 
(oblate here, cf. Fig.~\ref{fig:Tauvsgamma}b, and prolate in C\&H+92). 

\subsection{Logarithmic derivative~:  $\gamma = 1$} 
One reason why a global function does a poor job of fitting the mass distribution  is the irregular, noisy profiling outside (about)  the 
 half-mass radius. We illustrate this with a series of graphs of the logarithmic derivative of the density as function 
of radius. A piecewise linear fit to the  finite-difference scheme was done, such that 

\begin{equation} \left.\frac{d\log\rho(r)}{d\log r}\right|_i = \frac{ \log ( \rho_i/\rho_{i+k} ) }{\log (r_i/r_{i+k}) } 
\label{eq:dlogrho}  \end{equation} 
where $i := [1,1000-k]$ is the index of a mass shell, and $k \approx 30$ is constant. 
The value of $k$ is chosen so as  to sample a sufficiently large radial increment and avoid round-off 
errors in (\ref{eq:dlogrho}). We found that $k$ could be increased to $\approx 200$ without affecting the overall 
picture much. We then used a least-square fitting routine (Press et al. 1992) to find the best linear fit over $k$ 
points. The result was centered on the mean radius in the interval  $i, i+k$. 

On Fig.~\ref{fig:dlogHernfit}b we graph this derivative for the run $\gamma = 1$ listed in Table~\ref{tab:Gamma}. We have indicated the radii of the inner 1\%, 10\% and 20\% mass shells with vertical ticks on the 
figure. It is clear that the derivative is smooth and well fitted up to the 20\% mass shell; however beyond that
point large fluctuations are seen which indicate grainy or irregular substructures. Note also that the data match very well 
the logarithmic derivative of (\ref{eq:Hernquist}), given by a solid line on the 
figure, from the 
innermost 0.1\% mass bin up to roughly the 20\% mass radius. Note the apparent clustering 
around $d\log\rho/d\log r \approx -2.2$ also  between the 20\% and half-mass shells $(r_h \simeq 0.53)$.
 Beyond the half-mass radius the 
 derivative oscillates wildly between values of $-3 $ and $-4$; beyond $r = 1$ or so the data is more erratic. 

\subsection{Logarithmic derivative for all $\gamma's$}
It may come as a disappointment that the logarithmic derivative on Fig.~\ref{fig:dlogHernfit}b has 
not converged to  the anticipated constant value $= -1$, despite the very small innermost mass fraction (0.1\%) sampled.  We asked whether the same was true 
of all the runs with different $\gamma$'s. We therefore repeated the procedure for all runs listed in Table~\ref{tab:Gamma}, except for the case of $\gamma = 3/2$ where we took the results from the 800,000 particle run. 
The results are displayed on Fig.~\ref{fig:DlogRho}. Here again the vertical ticks indicate 1\%, 10\% and 20\% 
mass shells. For small values of $\gamma$, $d\log\rho/d\log r$ does not converge to a flat value as we approach the center. However, as we consider larger values of $\gamma$ we find hints of such a convergence, particularly for $\gamma \ge 3/2$.  
 Once more we find a suggestion of  a leveling off  around $d\log\rho/d\log r \approx -2.2$ 
at radii close to the 20\% mass shell for all the cases with $\gamma \ge 1$. 
 In their treatment of the  Boltzmann equation in one dimension, Hozumi et al. (2000) also 
found a power-law fit of index $\approx -2.1$ around the half-mass radius of their systems. 

\subsection{Circular velocity} 
\begin{figure}
\begin{center} 
\setlength{\unitlength}{1cm} 
\begin{picture}(8,8.5)(0,0){ 
	\put(.75,1.){ \epsfysize=0.5\textwidth 
		    \epsffile[ 100 150 450 600]
		   {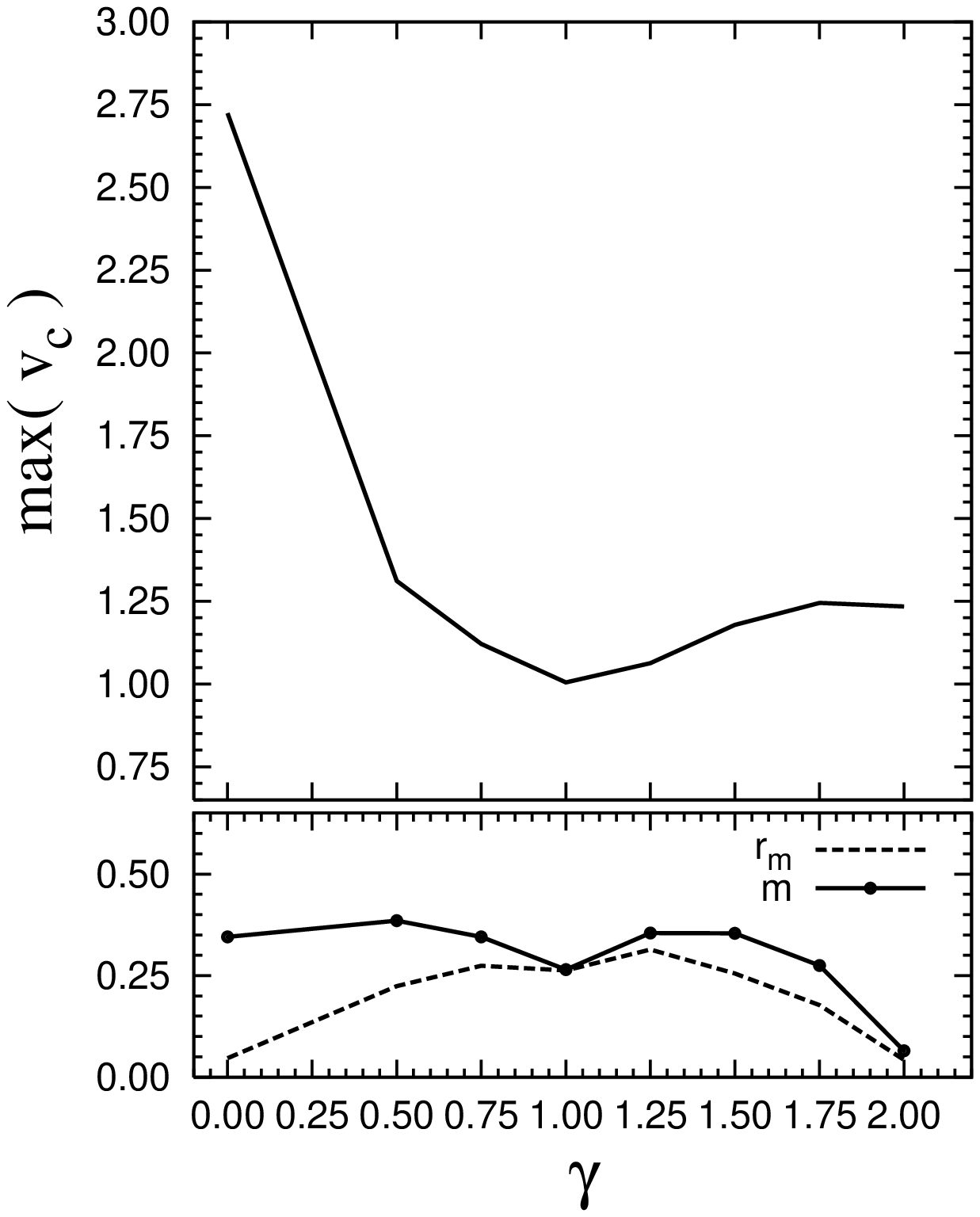} }
			}
\end{picture} 
      \caption{Maximum of the circular velocity $v_c$ as function of the initial power-index $\gamma$ (top panel). 
The lower panel shows the mass fraction ($m$) and radius ($r_m$) at maximum. }
         \label{fig:Maxvc}
\end{center}    \end{figure}

 On Fig.~\ref{fig:vc} 
we graph the (less noisy) circular velocity $v_c = \sqrt{G M / r} $ as function of the radius and of the integrated mass.  Note that $v_c \rightarrow $ constant when $\rho (r) \propto r^{-2}$. 
On the figure we also added a straight line 
($\propto r$ or $\propto M^{1/3}$) that $v_c$ would follow 
if the density flattened out at the centre. For our simulations, $v_c$ rises slowly as a function of radius and we find $v_c \approx r^{0.75}$ gives a rough fit for runs 
with $\gamma < 1$. Starting with the curve $\gamma = 1$, one can see a plateau appearing at a mass 
fraction $M(<r)/M \approx 0.1$, which becomes more prominent for increasing $\gamma$. 
The circular velocity 
drops off in a Keplerian tail at large radii, in all the cases. We remark that the
maximum of $v_c$ nearly always occurs at the mass fraction $M(<r)/M \approx 1/3$ while the maximum itself is a non-monotonic but nearly flat 
function of $\gamma$ (save $\gamma =0$, see Fig.~\ref{fig:Maxvc}). 
Therefore, while none of the equilibria can be mapped into another (different rotation curves, mass profiles), the results of collapse calculations for different  $\gamma$'s (and hence the accretion rates or history) 
 are fairly similar  in terms of max($v_c$), and of corresponding mass fraction and radius, a situation that parallels 
current debates about the universality of  resolved circular velocities and virial masses in cosmology (see e.g. Power et al. 2003; Navarro et al. 2004). Only the two cases of $\gamma = 0$ and 2 stand out.  These cases are particularly interesting, since they produce the largest and 
second largest maximum $v_c$, but the smallest corresponding radius and mass fraction (bottom panel, Fig.~\ref{fig:Maxvc}). 
\newline 

\subsection{Power-law limit} 
\begin{figure*}
\setlength{\unitlength}{2cm} 
\begin{picture}(6,3)(0,0) 
	\put(-1.2,1.){ \epsfysize=3.20in
		    \epsffile[ 100 150 450 600]
		   {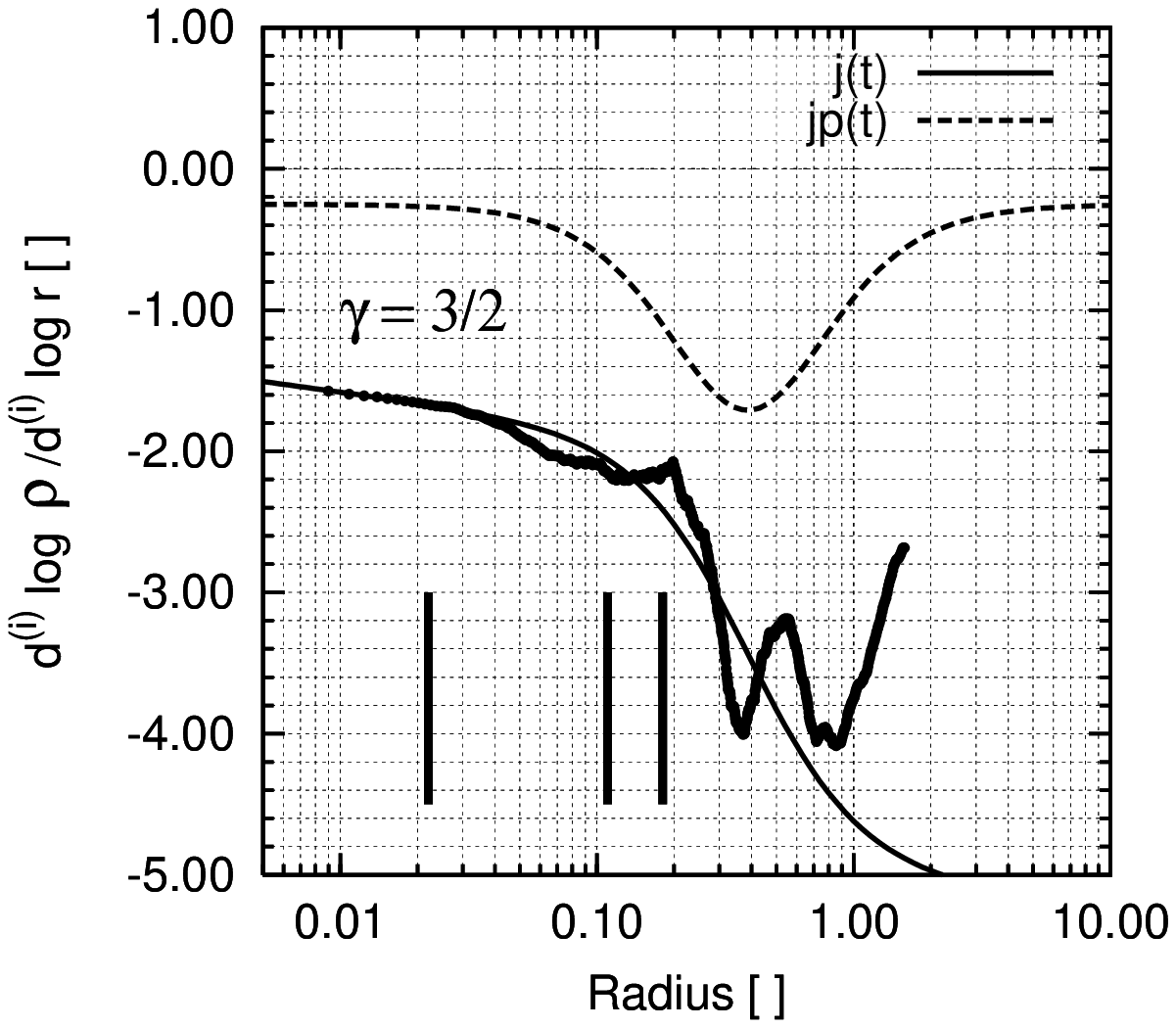} 
			}
	\put(1.8,1.){ \epsfysize=3.0in
		    \epsffile[ 100 150 450 600]
		   {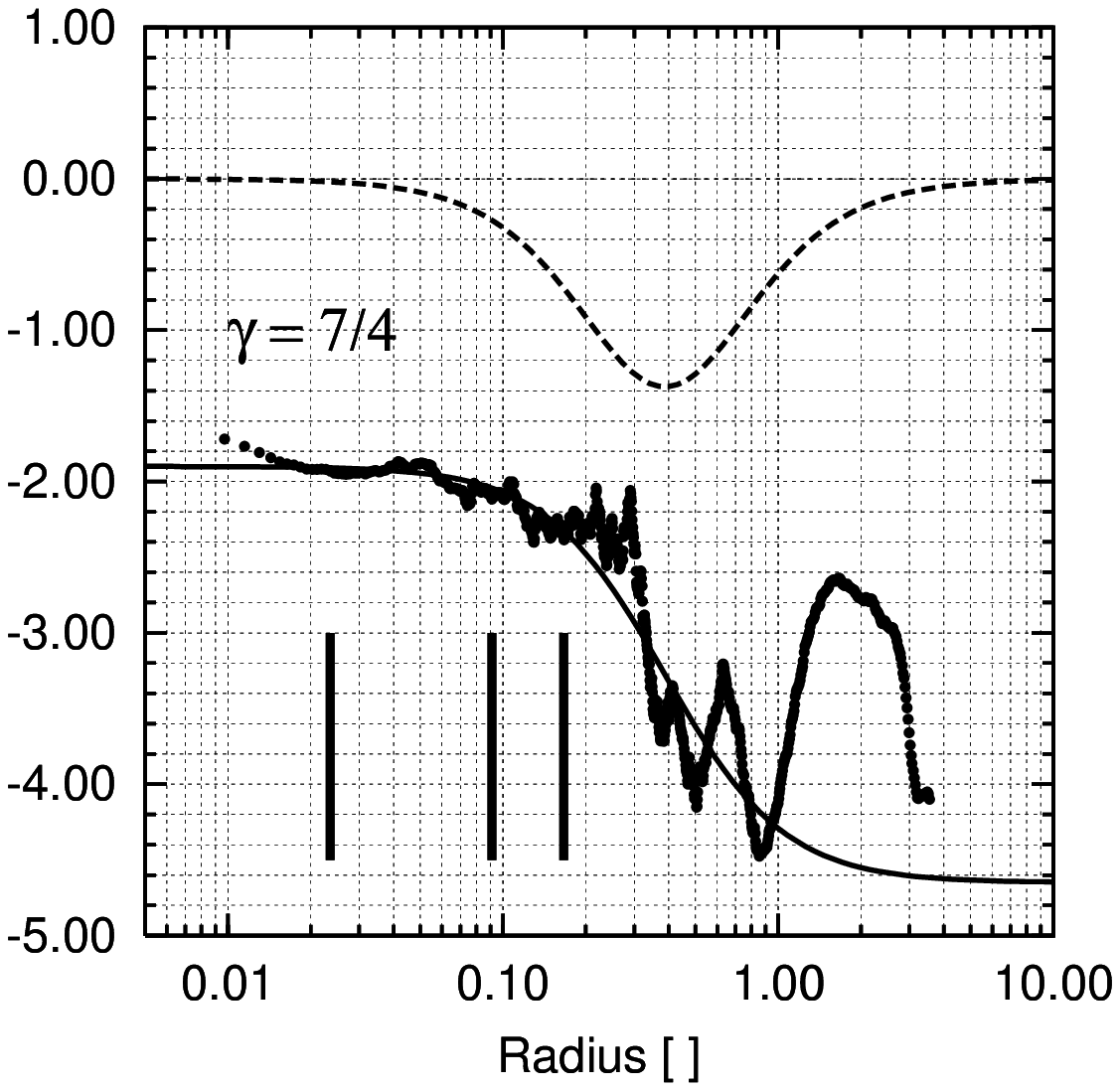} 
			}
	\put(4.65,1.){ \epsfysize=3.0in
		    \epsffile[ 100 150 450 600]
		   {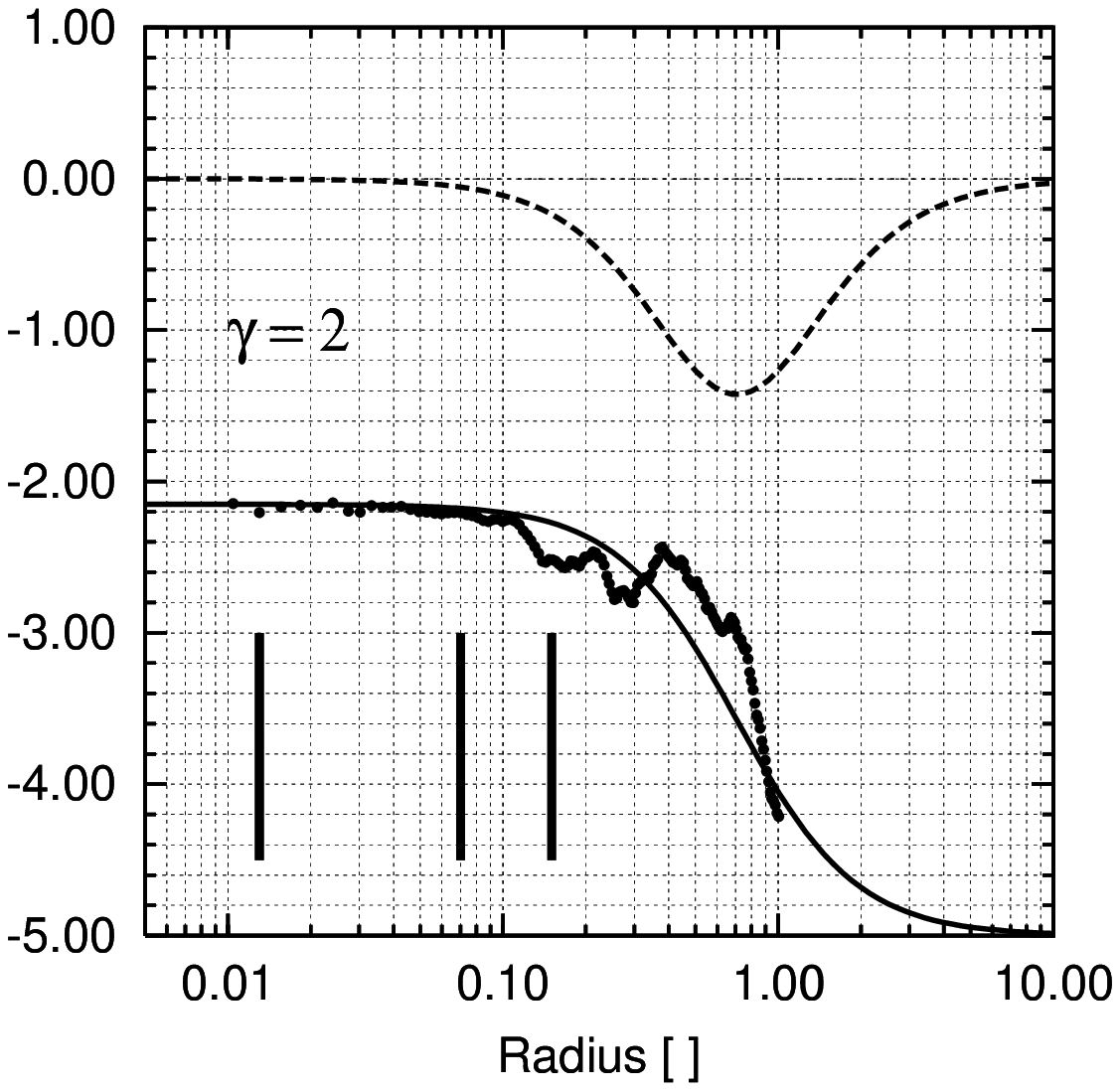} 
			}
\end{picture} 
      \caption{Logarithmic first and second 
derivatives versus radius for the three cases with 
the largest initial power index $\gamma$. The solid line is a fit  $(j)$ and the dashed 
line the derivative  of that fit ($j^\prime$, top-most curve). A strict power-law is ruled
out in the case $\gamma = 3/2$ everywhere. 
For $\gamma = 7/4$ a power-law inner profile 
out to $\approx $ a few percent of the mass is representative; while 
for $\gamma = 2$ the constant power-law region reaches $\approx$ 10\% of the total mass.}
\label{fig:fitDlogRho} 
   \end{figure*}

The search for a flat logarithmic derivative can be done through a functional fit which admits a linear regime in 
$\log r$. We modified (\ref{eq:Dehnen}) slightly to look for fits to the derivatives of the form 

\begin{equation} j(r) \equiv \frac{d\log\rho}{d\log r}  = -\gamma_f - \gamma_2 \log r - \beta \frac{r^2}{r^2 + r_o^2}  
\label{eq:dlogfit} 
\end{equation} 
where $r_0$ and $\beta$ are as before, the symbol $\gamma_f$ is used to distinguish the final and initial values of $\gamma$ and the constant $\gamma_2$ is equal to the second derivative 
at small radii to $O(r^2)$. We imposed $\beta = 4 - \gamma_f$ so three parameters remain free. 
A strict power-law at short distances must yield $\gamma_2 = 0$. We applied the 
functional $j(r)$ to three cases with $\gamma = 3/2, 7/4$ and $2$. The results are shown on Fig.~\ref{fig:fitDlogRho}. 
The data points are those of Fig.~\ref{fig:DlogRho} in each case, to which we have added $j(r)$ (solid line) and 
$j^\prime(r) = d j(r)/d\log r$ (dashed line). We find $j^\prime(r) < 0 $ everywhere for $\gamma = 3/2$, however a 
fit with $\gamma_2 = 0$ gave sensible results for both $\gamma = 7/4$ and 2. Even so the power-law regime in the  
case $\gamma = 7/4$ does not cover more than a few percent of the system mass, while it extends roughly  
out to 10\% of the mass for $\gamma =2 $.


Although the anticipated relation $\gamma_f = \lim_{r\rightarrow 0} - d\log\rho/d\log r \approx  \gamma =$ constant  
at small radii is not recovered, 
a strong correlation is found between $\gamma$ and the minimum of $-d\log\rho/d\log r$ read off Fig.~\ref{fig:DlogRho}. 
We graph $-d\log\rho/d\log r$ 
 for the innermost 1\% (solid) and 0.1\% (dotted) mass shells on Fig.~\ref{fig:maxdlogrho}. The diagonal on that figure
is  the sought equality $\gamma_f = \gamma$. We find a clear trend, 
 especially for $\gamma \ge 3/2$, the strongest for the innermost 
0.1\% of the mass. For runs with $\gamma < 1 $ the correlation is more suggestive, 
even for the innermost 0.1\% mass shell. 

 As long as we are concerned with the radial profile of the systems at small $r$, 
 these results show that memory of the initial configuration, while still there,  involves only a very small fraction of the total mass.
Correlations in binding energy have been noted by several authors 
 (see e.g. van Albada 1982; Henriksen \& Widrow 1999), so the  initially 
most-bound particles are those contributing to correlations seen on Fig.~\ref{fig:maxdlogrho}. 
 We  deduce   that  well-resolved  equilibria 
  could not have inner cusps with a logarithmic slope lower than $\gamma$ of the initial conditions. 
But significant deviations already at the 1\% mass fraction means that this conclusion 
 effectively applies to a very small volume. We may ask whether the velocity field also shows traces 
of the collapse, and in what measure. 

\section{Phase-mixing and relaxation}
\subsection{Cold initial conditions} 
Having brought to light a trend for the run of density in equilibrium as function of the 
initial power index $\gamma$ (Fig.~\ref{fig:maxdlogrho}), we now ask whether a 
similar relation exists for the velocity field. Repeated exchanges of 
kinetic energy modify  a particle's velocity by $\delta\boldv{v}$ (say) each time. When these boosts in 
velocity are randomly oriented the ensuing velocity vector is uncorrelated with the initial $\boldv{v}$. This will be true in an 
average sense when the sum of all perturbations exceeds the norm of the 
velocity vector, i.e. $\sum_i \delta\boldv{v}^2_i >  \boldv{v}^2$. 
 For systems collapsing from rest, $\boldv{v} = 0$, this condition 
 will be met first by particles that do not acquire large velocities 
 while the background potential changes rapidly. From these considerations,  
we would expect  the velocity field to be the one arising from violent relaxation for all orbits with relatively 
small in-fall velocities. Because of the enforced spherical symmetry 
of the initial conditions, all orbits are  radial 
($\boldv{v} = v_r \boldv{\hat{r}}$) at the on-set of collapse. The work by the gravitational 
force $\delta v_r^2 \propto r \nabla_r\phi \propto r^{2-\gamma} \rightarrow 0$ for all $\gamma$'s $< 2$. The kinetic energy remains small for all particles orbiting near the centre 
and hence small fluctuations in energy will wipe out memory of the coherent radial collapse.  

 A Maxwellian coarse-grained velocity distribution function (d.f.) 
 (more precisely: a sum of Maxwellians) is the signature of violent relaxation, a fully stochastic  
redistribution of kinetic energy. Merral \& Henriksen (2003) and Iguchi  et al. (2005) have obtained 
good agreement with this prediction  by 
integrating the collisions-less Boltzmann equation directly,  and evaluating  the velocity d.f. $f(v)$ 
in equilibrium at the origin of the coordinates. 
The same procedure can not be done here due to finite 
resolution. Instead,  
 we evaluated $f(v)$   for an  100,000-particle  $\gamma = 3/2$ calculation 
by sorting and binning in velocity space all particles within a given mass fraction. We picked a fraction $M(<r)/M = 2.5\%$ 
which samples the regime where the logarithmic derivative is (roughly) linear with the logarithm of the radius, that is, the density $\rho \propto r^{k \ln r}$ with $k$ some numerical constant (cf. Fig.\ref{fig:DlogRho}) and much larger than the softening length $\epsilon \simeq 0.002$. 
For comparison, we  also computed $f(v)$ for 
 all particles, taking care to normalise the d.f. so that it integrates to unity in each case.  

\begin{figure*}
\setlength{\unitlength}{1cm} 
\begin{picture}(8,9.5)(0,0) 

	\put(-3.5,2.5){ \epsfysize=0.58\textwidth
		    \epsffile[ 100 150 450 600]
		   {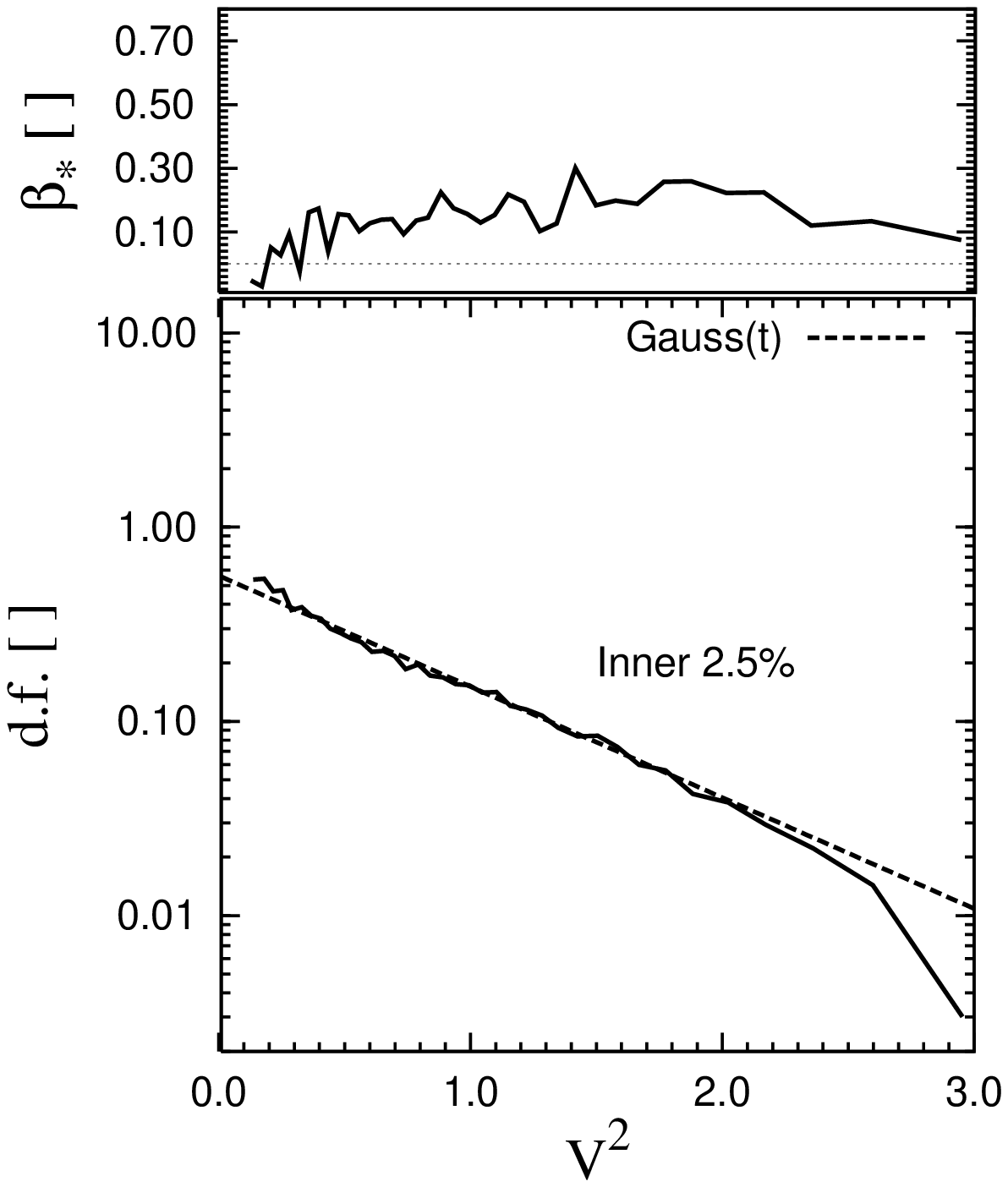}  
}
	\put(5.,2.5){ \epsfysize=0.6\textwidth 
		    \epsffile[ 100 150 450 600]
		   {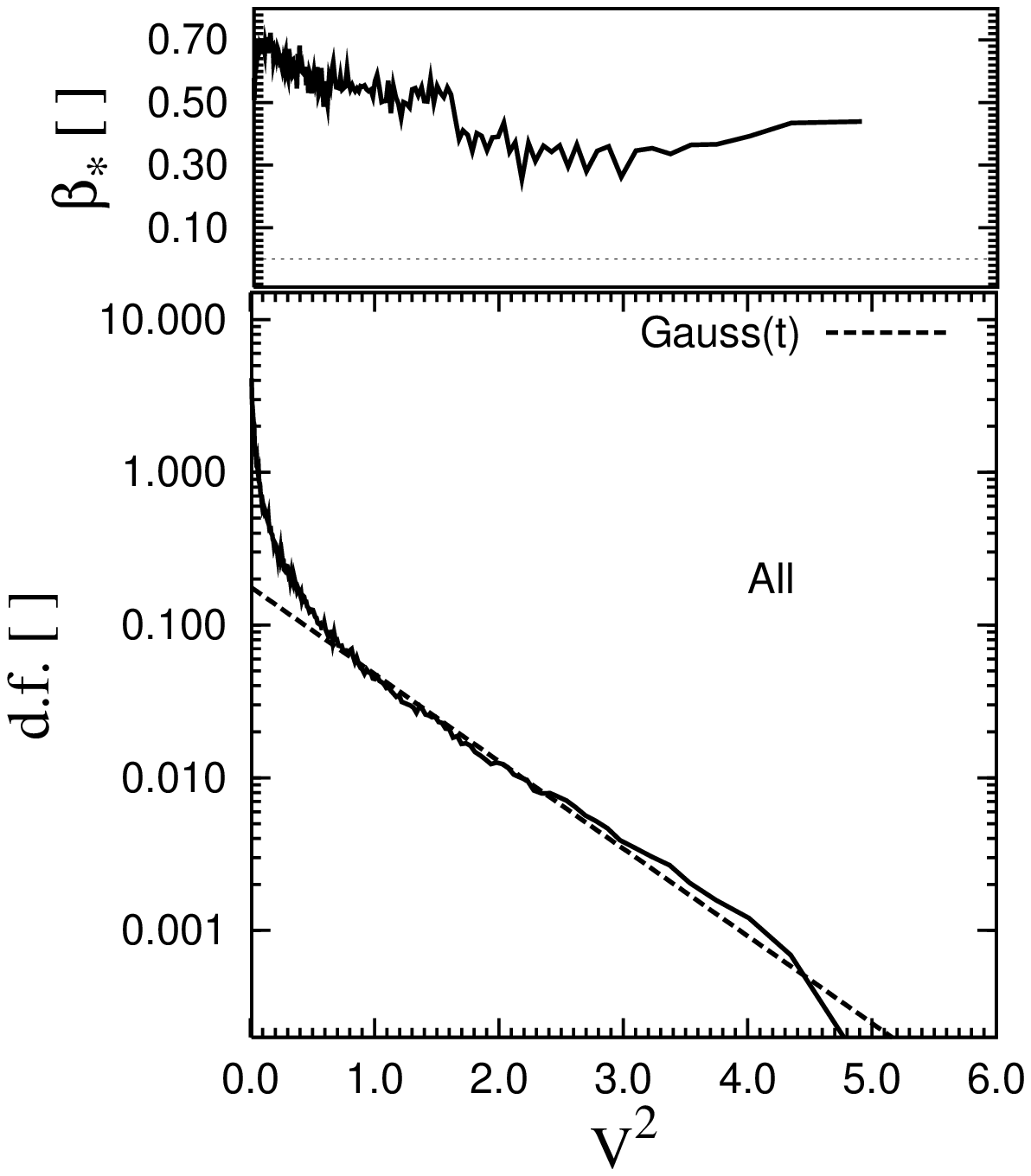} 
                  }
\end{picture} 
      \caption{The velocity d.f. $f(v)$ as function of $v^2$. Note the logarithmic scale. The function was 
constructed for the innermost 2.5\% particles at the end of the simulation (left-hand panel). The same quantity 
computed for the system as a whole is shown for comparison (right-hand panel). The dashed curve is the same Maxwellian on 
both panels; the axes were rescaled for better display. The top panels show runs of the anisotropy parameter $\beta_\ast \equiv 1 - 0.5\, \langle\sigma^2_\perp/\sigma^2_r\rangle$; $\beta_\ast > 0$ for systems with excess radial motion. The inner part (left-hand set) is more isotropic, while 
globally the system shows an  increasingly anisotropic velocity field as we reach the edge of the system ($v^2\rightarrow 0$).} 
         \label{fig:dfv}
   \end{figure*}

\begin{figure}
\begin{center}
\begin{picture}(200,200){ 
\put(-17.,-35.){\epsfig{file=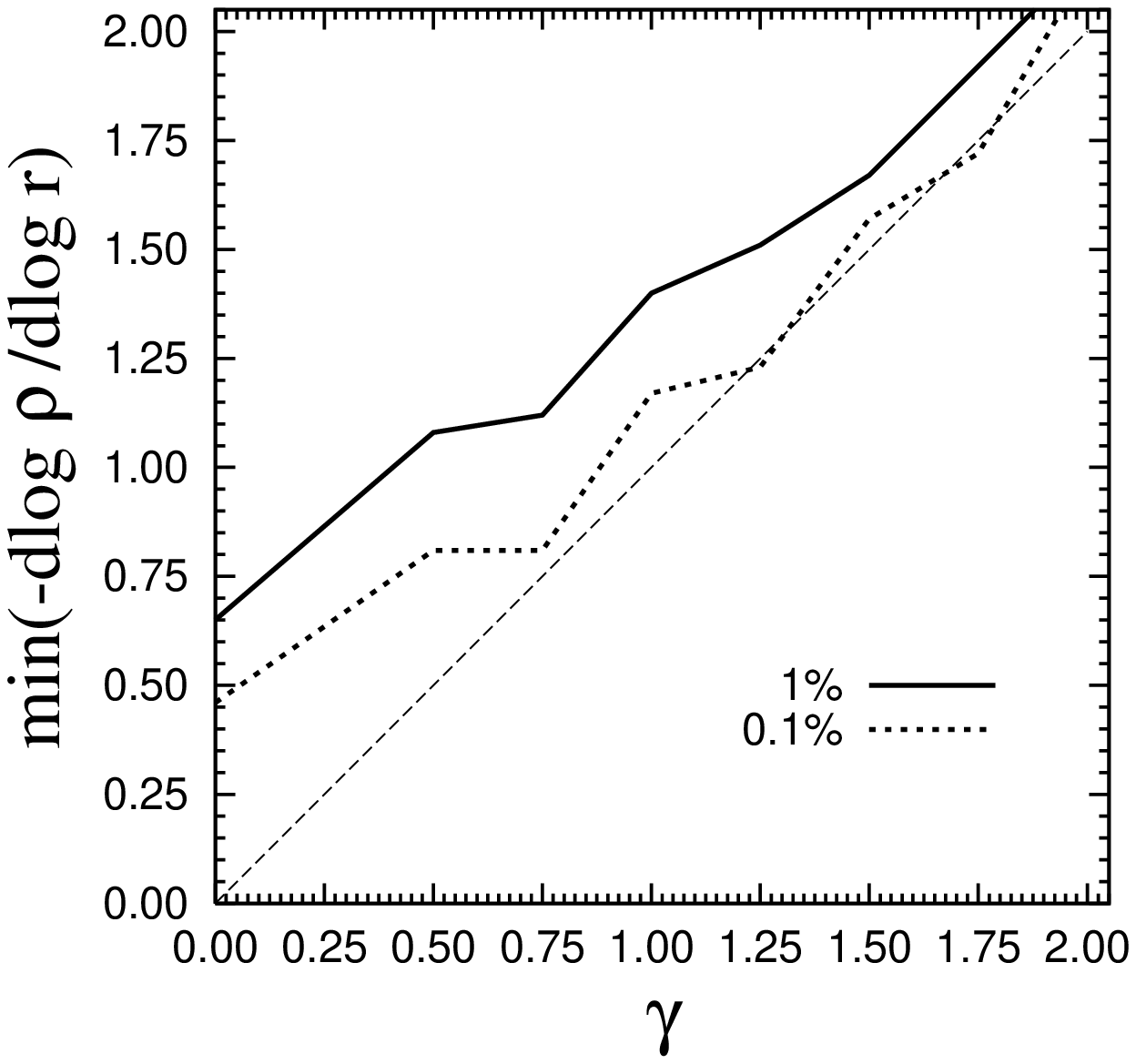, width=0.4\textwidth}}
}\end{picture} 
 \caption{Value of the logarithmic derivative $- d\log\rho/d\log r$ as function of the initial conditions 
 power index $\gamma$ for two mass shells of 1\% (solid line) and 0.1\% (dash). The latter is the minimum positive 
 value obtained for  each case. The diagonal broken line is equality.} 
\label{fig:maxdlogrho}
\end{center}
\end{figure}

The results are displayed on Fig.\ref{fig:dfv}. We find excellent agreement with a Maxwellian 
profile for the inner 2.5\% mass 
sample despite the rough dataset used (2,500 particles only)\footnote{A repeat with the 800,000-particle run, or 
by averaging over  time intervals $\gg $ the local dynamical time, only confirms the quality of the agreement.}. 
 Applied to the system as a whole, we find an overabundance  
of low-velocity particles, while overall the fit is not as good (see Fig.~\ref{fig:dfv}, right-hand panel). 
The excess probability density at low-velocities with respect to an  isotropic Maxwellian velocity d.f. may be understood as an overabundance of stars on low-energy radial orbits. Stars on such orbits spend more time at apogalacticon, where their velocity is relatively low.
This is confirmed by looking at  runs of the isotropy parameter $\beta_\ast \equiv 1 - 1/2\, \langle \sigma^2_\perp/ \sigma^2_r\rangle $ distinguishing between perpendicular ($\perp$) and radial velocity components (top panels, Fig.~\ref{fig:dfv}). 
Clearly $\beta_\ast = 0 $ for an isotropic velocity field, and $\beta_\ast > 0$ when the system shows an excess of radial motion. We find the inner region more nearly isotropic (left-hand set), while globally $\beta_\ast$ assumes larger values and increases as $v^2$ decreases (right-hand set). 

In conclusion, we find a good fit to the inner region velocity d.f. with a Maxwellian, while globally this does not hold. 
These results go in the same directions as the findings by others (e.g., Funato et al. 1992; Merrall \& Henriksen 2003; 
see also Bertin \& Trenti 2003) but for fully three-dimensional calculations. By the time particles fall in with large velocities, much orbit crossing (and therefore
phase mixing) has gone on in the centre. As a result, the potential varies more smoothly in time and these particles 
preserve significant radial motion. This is seen in animated snapshots, when 
the last mass shell has collapsed and particles rebound from the central region. 
A fraction of them are unbound and leave the system.


\subsection{Warm initial conditions} 
 We wish to assess how equilibrium profiles of 
 simulations  with $Q \ne 0 $ initially differ 
 from those starting from rest. Since kinetic energy provides pressure
 support, a run with $Q > 0$ will collapse more slowly (reduced mass in-fall, $\dot{m}(t)$). 
 The expectation drawn from simulations with different $\gamma$'s 
 is that a reduced rate of in-fall (large $\gamma$, cf. Eq. \ref{eq:mdot}) 
   would favour   oblate morphology in equilibrium (cf. Fig.~\ref{fig:Tauvsgamma}b). 
  A\&M+90 have shown that sufficiently warm collapse simulations 
 shut off ROI's and preserve the symmetry of the initial configuration.  (The limiting case of an initial $Q = 1$ equilibrium 
  must trivially preserve spherical symmetry.)  
  The question  at stake here is whether ROI modes of instability 
  are washed out progressively  as $Q$  increases from zero, 
  leading to more and more spherically symmetric equilibria;  
  or whether more oblate equilibria first develop for $Q > 0$ but small, 
  before being washed out.
  
In order to answer this question we ran a series of simulations with warm 
initial conditions. We set up Dehnen (1993) density profiles, Eq. (\ref{eq:Dehnen}),  
with $\gamma  = 3/2, \beta = 5/2$, $r_0 = 0.756..$, 
which are truncated at  radius $r_{t} \approx 3.76$ so that the density profiles 
of these models match the density of the power-law models at the centre.  
This choice of scales  allowed us to use the same numerical setup (smoothing length and 
time-steps) as for the power-law computations since the dynamics at the centre is 
identical in both cases.  The total mass  $M ( < r_t) = 1$ as before.  
As explained in \S2, we attributed velocities according to 
an isotropic  Maxwellian velocity d.f. constrained to satisfy locally 
the first moments of the Jeans equations (Binney \& Tremaine 1987).  
The outcome of such studies  are not sensitive to  details of the velocity field 
but to the global value of $Q$ (Barnes et al. 1986; A\&M+90).

\begin{figure*}
\begin{center}
\begin{picture}(400,400){ 
}\end{picture} 
 \caption{Phase-space sections showing the radial velocity as function of 
 radius during in-fall for two runs, one starting from zero velocities (d002, left-hand set) 
 and the other from sub-virial velocities (h002, right-hand set).  A similarity pattern 
 shows up in the form of concentric rings, which are progressively deleted at later 
 times.} 
\label{fig:self-similar}
\end{center}
\end{figure*} 

\subsubsection{Similarity patterns} 
A self-gravitating collapse proceeding from power-law initial conditions 
soon develops single-valued orbital patterns, such that any star follows  
the same orbit as any other once the scales of mass and lengths are 
renormalised under a time-transformation. The trajectories of the stars map out 
a unique path and the flow is said to be self-similar. The growth of similarity patterns 
saturates to give way to a phase-mixing instability leading to virialised equilibrium (Henriksen \& Widrow 
1997, 1999). An example of this process is displayed on 
Fig.~\ref{fig:self-similar}, which compares the run of the radial velocity with radius for two time 
sequences, one corresponding to a cold collapse (d002) and the other to a warm one (h002).
 From this, and other similar plots for other simulations and other
times (not shown here), it is
clear  that the self-similar pattern remains well  
defined for a longer time in the warm run
than in the cold one. This is in good agreement with Merral \&
Henriksen (2003), who argued that warm initial  
conditions  slow down the transition to a phase-mixing
instability. This trend, however, is not monotonic. Indeed, for very hot
runs, such  similarity patterns hardly show up. Thus, intermediate
values of $Q$, of the order of 0.1 or 0.2, are the optimum for their 
formation and maintenance. 

We also note that, for the warm simulation, the radial velocity dispersion is
relatively large at the earliest time, and for sufficiently high $Q$
values some particles even have $v_r > 0$ and outward motion. 
As time runs, the radial flow becomes cooler (the one-stream at large
radii becomes thinner). In the late stages the  
one-stream in-falling material is more and more fine-tuned
(single-valued), as in  the  
cold collapse. This observation will serve us below when interpreting 
 the structural evolution of the models during collapse. 

\subsubsection{Oblate, and then not} 
 Table~\ref{tab:Hernquist} lists the overall morphology in equilibrium 
  of a series of Dehnen models with  $\gamma = 3/2$ and different  $Q$ initially.  
 The total runtime was 80 time 
units in each case. All collapses with $Q > 0.4$ (not listed in Table~\ref{tab:Hernquist}) led to  highly spherical equilibria. For $Q \le 0.4$ we get either prolate or oblate  morphologies. For the coldest cases ($Q \le 0.05$) the equilibria are oblate, as the scale-free models. With increasing $Q$, the equilibria first shift to prolate and then to spherical symmetry. This is rather different from the monotonic trend we would 
 have anticipated, had the aspherical modes of instability been washed out progressively 
 with increasing $Q$. The transition is very sharp around $Q \approx 1/10$. Two possible 
 explanations
of this phenomenon  are on hand. The first invokes the growth of a bending-mode of instability, similarly to self-gravitating thin 
 discs (see Merritt 1999). Here, the  highly elongated prolate morphology of the cold (low $Q$'s) 
 runs implies motion mainly down the semi-major axis. 
 As $Q$ is reduced and the bar achieves a highly prolate shape, an off-axis mode of instability will develop owing to 
 the large velocities of the particles parallel to the semi-major axis. Comparing centrifugal and restoring forces 
 in the case of a thin bar distorted by a sinusoidal bend of wave-number $\kappa = 2\pi/\lambda$
 and amplitude $\cal A$ immediately gives the condition $ {\cal A}\kappa > $  constant 
 for growth of the mode (the precise value of the constant  is of no interest here). 
 A similar relation applies for discs. 
 Clumping modes of short wavelengths (large $\kappa$) will develop 
 more easily in cooler systems, whence the observed transition to oblate morphology. 
 An alternative explanation is that   a two-stream 
 mode of instability develops more fully in the deep potential well of the colder collapses, 
 when the in-falling material reaches larger velocities and more orbit-crossing takes place at the centre.  Clearly a stability analysis well into 
 the non-linear regime is required to determine which one of the two (or other) 
 types of instability  prevails.  Such an analysis goes far beyond the objectives 
 of the present paper and will not be attempted here.  
 The close link between equilibrium properties and 
initial virial ratio was noted long ago for spherically averaged values (van Albada 1982; Mclynn 1984; A\&M+90), however, to the best of our knowledge, 
the transition from oblate to prolate morphology and then to 
spherical symmetry as $Q$ increases has not been stressed before. 
 \newline

   We close this section with a comparison of the cold $Q = 0$ collapse from a power-law 
 profile with the corresponding Dehnen model. 
The cold Dehnen run listed in Table~\ref{tab:Hernquist} 
reached $\tau_\ast \approx 0.4$ in equilibrium, i.e. more oblate  than what was obtained from an $\gamma = 3/2$ power-law initial conditions (cf. Fig.~\ref{fig:Tau} and Table~\ref{tab:Gamma}). 
  Since the mass profile of a Dehnen model is steeper  at 
  large radii than $\rho \propto r^{-3/2}$, mass shells falling in from large 
  $r$ take longer to reach the origin in that case. In other words, the rate of mass in-fall 
  of the Dehnen  sphere is lower compared to the case of collapse from a power-law profile. This further supports  
  our claim that a reduced rate of mass in-fall $\dot{m}$ (cf. Eq.~\ref{eq:mdot}) leads to more oblate equilibria, at least when the initial virial ratio $Q = 0 $.  

\section{Summary and discussion} 
Using $N$-body computer simulations, we have shown that collapsing self-gravitating spheres
 develop oblate or prolate triaxial figures of equilibrium. In a study of numerical 
convergence, we have shown that 
an insufficient number of particles $N$, or  linear resolution $\epsilon$, can give wrong results for the morphology. As $N$ increases 
we find evidence for a slow drift from prolate to oblate morphology; the trend is similar but more 
pronounced when we {\it reduce} the smoothing length $\epsilon$. 
Previous studies had found prominently prolate structures of equilibrium, but this appears to have 
been due partly to  the numerical setup used. The degree of symmetry of the 
virialised structures is highly sensitive to the phase-mixing which takes place during in-fall. 
 We used scale-free power-law initial density profiles and found that steeper powers lead to more 
oblate equilibria. We also noted that increasing the initial virial ratio $Q > 0$ leads to  
prolate equilibria. Increasing $Q$  to yet larger values 
has the effect of maintaining spherical symmetry, by 
shutting down aspherical modes of instability (Barnes et al. 1986). For very low-$Q$ calculations, 
we noted that the prolate equilibria give  way to oblate shape and have invoked two mechanisms 
that likely play a role in this transition (\S6.2).  A full analysis of that phenomenon 
is deferred to a future study. 

All our simulations lead to peaked central regions. When fitting the run of spherically-averaged
volume density with radius we found that the logarithmic derivative of the profiles converges
only slowly to a power-law, at least when $\gamma \le 3/2$ (Fig.~\ref{fig:DlogRho}). 
The logarithmic derivative in the central region of the equilibrium systems is well 
correlated with $\gamma$ when $\gamma \ge 3/2$ (Fig.~\ref{fig:maxdlogrho}); in all 
cases, the central cusps would include only a very small fraction of the total mass, 
on a scale where the impact of baryonic (i.e., gas) physics would not be negligible. 
 Collisional relaxation effects are always a worry in  dense central regions. 
  We checked that the collisional time of our simulations is too long, even at the 
centre, for two-body encounters to play a role.  Power et al. (2003)  conducted a 
thorough analysis of these effects. We find our simulations on the safe side of their 
resolution criterion (see their Fig.~14). This boosts our confidence that the properties 
of the cusps and the evolution we observed at the innermost 20\% mass shell is 
a genuine effect of the complex triaxial structure of the equilibria, 
and not a numerical artifact due to two-body relaxation. 

We also found that  steep initial density profiles lead to equilibria  with a 
 leveling off in the circular velocity as function of radius. This suggests a run of 
density in equilibrium $\rho \propto r^{-2}$ for a non-negligible fraction of the mass. 
This feature was also reported by Hozumi et al. (2000), who approached the problem from 
the angle of the collision-less Boltzmann equation. \newline 

We have observed evolution in time for the innermost region of the equilibrium profiles, 
such that the small-scale profile shifts away from triaxiality, and toward oblate 
axi-symmetry (see also Theis \& Spurzem 1999; Heller 1999).  This would certainly impact
 on the orbital structure of the equilibrium. In a follow-up investigation,  
we will explore the orbital structure of the equilibria obtained here in details to bridge over 
with models of triaxial ellipticals based on distribution functions. 

\section*{Acknowledgments}
It is a pleasure to thank A. Bosma for
useful and motivating discussions and Jean-Charles
Lambert for his invaluable help with the simulation software and the
administration of the runs. 
E.A.  also thanks the observatoire de Marseille, the region PACA, the
INSU/CNRS and the University of Aix-Marseille I for funds to develop
the computing facilities used for the calculations in this paper. This 
project benefitted from PNG grants  awarded in 2002, 2003 \& 2004 and 
from the SFB439 programme  in Heidelberg, Germany, when it 
was conceived in 2001. C.M.B. thanks Rainer Spurzem for support during
these early stages. 

 \label{lastpage}

\newpage 
\onecolumn 

\begin{table} 
\caption{List of runs with cold scale-free initial conditions. 
\label{tab:IC} }
\begin{center} 
\begin{tabular}{lllll|}  
Name & N& $\theta_c$ & 1/$\epsilon$ & $\gamma$ \\ 
           & [$10^3$]  &     &                      & \\
d001 & 100 & 0.7 & 512 & 3/2 \\ 
d004 & 100 & 0.6 & 512 & 3/2 \\ 
d003 & 100 & 0.5 & 512 & 3/2 \\ 
d002 & 100 & 0.4 & 512 & 3/2 \\ 
d005 & 100 & 0.3 & 512 & 3/2 \\  
d006 & 100 & 0.2 & 512 & 3/2 \\ 
\\
d019 & 10 & 0.4 & 512 & 3/2 \\ 
d010 & 25 & 0.4 & 512 & 3/2  \\ 
d008 & 32 & 0.4 & 512 & 3/2  \\ 
d009 & 50 & 0.4 & 512 & 3/2  \\ 
d020 & 100 & 0.4 & 512 & 3/2 \\ 
d022 & 200 & 0.4 & 512 & 3/2 \\ 
d023 & 400 & 0.4 & 512 & 3/2 \\ 
d007 & 800 & 0.4 & 512 & 3/2 \\ 
\\
d011 & 100 & 0.4 & 512 & 0   \\ 
d012 & 100 & 0.4 & 512 & 1/2 \\ 
d015 & 100 & 0.4 & 512 & 3/4  \\ 
d013 & 100 & 0.4 & 512 & 1  \\ 
d016 & 100 & 0.4 & 512 & 5/4  \\ 
d017 & 100 & 0.4 & 512 & 7/4  \\ 
d014 & 100 & 0.4 & 512 & 2  \\ 
\\
d031 & 100 & 0.4 & 32   & 3/2   \\ 
d030 & 100 & 0.4 & 64   & 3/2   \\ 
d025 & 100 & 0.4 & 128  & 3/2   \\ 
d024 & 100 & 0.4 & 256  & 3/2   \\ 
d021 & 100 & 0.4 & 512 & 3/2  \\ 
d026 & 100 & 0.4 & 1024 & 3/2   \\ 

\end{tabular}
\end{center} 

\end{table}

\begin{table} 
\caption{Morphology of equilibria obtained for runs with 
different particle number $N$. The smoothing is $\epsilon = 1/512$ and $\gamma = 3/2$ 
in all cases. Values are averages over the innermost 80\% system mass. 
\label{tab:N}  }
\begin{center} 
\begin{tabular}{lrllcc} 
Name  & $ N $ & $b/a$ & $c/a$ & $<\tau_\ast>$ \\ 
      & [$10^3$]\ & 
\\
d007  & 800 & 0.76 & 0.45 & 0.21 \\  
d023   &400 &  0.81 & 0.47  & 0.28 \\ 
d022   &200  &  0.76 & 0.47  & 0.20 \\ 
d002 & 100  &  0.78 & 0.45 & 0.27 \\ 
d009   &50  &  0.72 &  0.48 & 0.06 \\ 
d008   &32  &  0.77 &  0.52 &  0.12 \\ 
d010   &25  &  0.78 &  0.52 &  0.12 \\ 
d019  &10  &  0.80 &  0.57  &  0.10 \\ 
\end{tabular}
\end{center} 

\end{table}

\begin{table} 
\caption{Morphology of equilibria obtained for runs with 
different smoothing $\epsilon$. The number of particles is $N = 100,000$ and $\gamma = 3/2$ 
in all cases. Values are averages over the innermost 80\% system mass. 
\label{tab:Epsilon}  }
\begin{center} 
\begin{tabular}{lrllcccccc}  
Name  &  1/$\epsilon$ & $b/a$ & $c/a$ & $<\tau_\ast>$ \\
\\
d031 & 32  & 0.65 & 0.43   & -0.03 \\ 
d030 & 64  & 0.70 & 0.44   & 0.08  \\ 
d025 & 128 & 0.72 & 0.44   & 0.14  \\ 
d024 & 256 & 0.86 & 0.47   & 0.38  \\
d021 & 512 & 0.70 & 0.40   & 0.17  \\
d002 & 512 & 0.82 & 0.55   & 0.30  \\
d026 & 1024& 0.98 & 0.56   &  0.47 \\

\end{tabular}
\end{center} 

\end{table}

\begin{table} 
\caption{Morphology of equilibria obtained for runs with 
different initial power index $\gamma$. The smoothing is $\epsilon = 1/512$ 
and $N = 100,000$ 
in each case. Values are averages over the innermost 80\% system mass. 
\label{tab:Gamma}  }
\begin{center} 
\begin{tabular}{lcllcc} 
Name  & $ \gamma $ & $b/a$ & $c/a$ & $<\tau_\ast>$ \\ 
\\
d012 & 1/2  & 0.65  & 0.48  & -0.13  \\ 
d015 & 3/4  &  0.64 & 0.44  & +0.08 \\ 
d013 & 1    &  0.74 &  0.45 & +0.14 \\ 
d016 & 5/4  & 0.71  & 0.44 & +0.10 \\ 
d002 & 3/2  &  0.78 & 0.45 & +0.27 \\ 
d017 & 7/4  &  0.88 & 0.49 & +0.39 \\ 
d014 & 2  & 0.98  & 0.58    & +0.47 \\ 

\end{tabular}
\end{center} 

\end{table}
\newpage 

\begin{table} 
\caption{Parameters of the calculations based on $\gamma = 3/2$ Dehnen  models (Eq. 2).
 The smoothing length is $\epsilon = 1/512$ and $N$ = 100,000 in each case.\label{tab:Hernquist}}
\begin{center} 
\begin{tabular}{lrllccl} 
Name  & $Q$  & $b/a$ & $c/a$ & $<\tau_\ast>$ & Comment  
\\ \\
h001 & 0.0 & 0.98  & 0.61  & 0.42 & Highly oblate  \\ 
h009 & 0.05 & 0.63  & 0.52  & -0.18 &                       \\ 
h002 & 0.1 & 0.63  & 0.62   & -0.41  & Highly prolate \\ 
h008 & 0.15 & 0.74  & 0.73  & -0.21 &                         \\ 
h003 & 0.2 & 0.98 & 0.96  & 0.00  & (quasi-) spherical \\ 
h004 & 0.4 & $>0.98$ & $>0.98$   & 0.00   & spherical \\ 
\end{tabular}
\end{center} 
\label{tab:hot}
\end{table}

\end{document}